\input pictex.tex   

\immediate\write10{Package DCpic 2002/05/16 v4.0}

\catcode`!=11 

\newcount\aux%
\newcount\auxa%
\newcount\auxb%
\newcount\m%
\newcount\n%
\newcount\x%
\newcount\y%
\newcount\xl%
\newcount\yl%
\newcount\d%
\newcount\dnm%
\newcount\xa%
\newcount\xb%
\newcount\xmed%
\newcount\xc%
\newcount\xd%
\newcount\ya%
\newcount\yb%
\newcount\ymed%
\newcount\yc%
\newcount\yd
\newcount\expansao%
\newcount\tipografo
\newcount\distanciaobjmor
\newcount\tipoarco
\newif\ifpara%
\newbox\caixa%
\newbox\caixaaux%
\newif\ifnvazia%
\newif\ifvazia%
\newif\ifcompara%
\newif\ifdiferentes%
\newcount\xaux%
\newcount\yaux%
\newcount\guardaauxa%
\newcount\alt%
\newcount\larg%
\newcount\prof%
\newcount\auxqx
\newcount\auxqy
\newif\ifajusta%
\newif\ifajustadist
\def\objPartida{}%
\def\objChegada{}%
\def\objNulo{}%


\def\!vazia{:}

\def\!pilhanvazia#1{\let\arg=#1%
\if:\arg\ \nvaziafalse\vaziatrue \else \nvaziatrue\vaziafalse\fi}

\def\!coloca#1#2{\edef\pilha{#1.#2}}

\def\!guarda(#1)(#2,#3)(#4,#5,#6){\def\id{#1}%
\xaux=#2%
\yaux=#3%
\alt=#4%
\larg=#5%
\prof=#6%
}

\def\!topaux#1.#2:{\!guarda#1}
\def\!topo#1{\expandafter\!topaux#1}

\def\!popaux#1.#2:{\def\pilha{#2:}}
\def\!retira#1{\expandafter\!popaux#1}

\def\!comparaaux#1#2{\let\argA=#1\let\argB=#2%
\ifx\argA\argB\comparatrue\diferentesfalse\else\comparafalse\diferentestrue\fi}

\def\!compara#1#2{\!comparaaux{#1}{#2}}

\def\!absoluto#1#2{\n=#1%
  \ifnum \n > 0
    #2=\n
  \else
    \multiply \n by -1
    #2=\n
  \fi}




\def\commdiag{0}


\def\!ajusta#1#2#3#4#5#6{\aux=#5%
  \let\auxobj=#6%
  \ifcase \tipografo    
    \ifnum\number\aux=10 
      \ajustadisttrue 
    \else
      \ajustadistfalse  
    \fi
  \else  
   \ajustadistfalse
  \fi
  \ifajustadist
   \let\pilhaaux=\pilha%
   \loop%
     \!topo{\pilha}%
     \!retira{\pilha}%
     \!compara{\id}{\auxobj}%
     \ifcompara\nvaziafalse \else\!pilhanvazia\pilha \fi%
     \ifnvazia%
   \repeat%
   \let\pilha=\pilhaaux%
   \ifvazia%
    \ifdiferentes%
     \larg=1310720
     \prof=655360%
     \alt=655360%
    \fi%
   \fi%
   \divide\larg by 131072
   \divide\prof by 65536
   \divide\alt by 65536
   \ifnum\number\y=\number\yl
    \advance\larg by 3
    \ifnum\number\larg>\aux
     #5=\larg
    \fi
   \else
    \ifnum\number\x=\number\xl
     \ifnum\number\yl>\number\y
      \ifnum\number\alt>\aux
       #5=\alt
      \fi
     \else
      \advance\prof by 5
      \ifnum\number\prof>\aux
       #5=\prof
      \fi
     \fi
    \else
     \auxqx=\x
     \advance\auxqx by -\xl
     \!absoluto{\auxqx}{\auxqx}%
     \auxqy=\y
     \advance\auxqy by -\yl
     \!absoluto{\auxqy}{\auxqy}%
     \ifnum\auxqx>\auxqy
      \ifnum\larg<10
       \larg=10
      \fi
      \advance\larg by 3
      #5=\larg
     \else
      \ifnum\yl>\y
       \ifnum\larg<10
        \larg=10
       \fi
      \advance\alt by 6
       #5=\alt
      \else
      \advance\prof by 11
       #5=\prof
      \fi
     \fi
    \fi
   \fi
\fi} 

\def\!raiz#1#2{\n=#1%
  \m=1%
  \loop
    \aux=\m%
    \advance \aux by 1%
    \multiply \aux by \aux%
    \ifnum \aux < \n%
      \advance \m by 1%
      \paratrue%
    \else\ifnum \aux=\n%
      \advance \m by 1%
      \paratrue%
       \else\parafalse%
       \fi
    \fi
  \ifpara%
  \repeat
#2=\m}

\def\!ucoord#1#2#3#4#5#6#7{\aux=#2%
  \advance \aux by -#1%
  \multiply \aux by #4%
  \divide \aux by #5%
  \ifnum #7 = -1 \multiply \aux by -1 \fi%
  \advance \aux by #3%
#6=\aux}

\def\!quadrado#1#2#3{\aux=#1%
  \advance \aux by -#2%
  \multiply \aux by \aux%
#3=\aux}

\def\!distnomemor#1#2#3#4#5#6{\setbox0=\hbox{#5}%
  \aux=#1
  \advance \aux by -#3
  \ifnum \aux=0
     \aux=\wd0 \divide \aux by 131072
     \advance \aux by 3
     #6=\aux
  \else
     \aux=#2
     \advance \aux by -#4
     \ifnum \aux=0
        \aux=\ht0 \advance \aux by \dp0 \divide \aux by 131072
        \advance \aux by 3
        #6=\aux%
     \else
     #6=3
     \fi
   \fi
}

\def\begindc#1{\!ifnextchar[{\!begindc{#1}}{\!begindc{#1}[30]}}
\def\!begindc#1[#2]{\beginpicture 
  \let\pilha=\!vazia
  \setcoordinatesystem units <1pt,1pt>
  \expansao=#2
  \ifcase #1
    \distanciaobjmor=10
    \tipoarco=0         
    \tipografo=0        
  \or
    \distanciaobjmor=2
    \tipoarco=0         
    \tipografo=1        
  \or
    \distanciaobjmor=1
    \tipoarco=2         
    \tipografo=2        
  \or
    \distanciaobjmor=8
    \tipoarco=0         
    \tipografo=3        
  \or
    \distanciaobjmor=8
    \tipoarco=2         
    \tipografo=4        
  \fi}

\def\enddc{\endpicture}

\def\mor{%
  \!ifnextchar({\!morxy}{\!morObjA}}
\def\!morxy(#1,#2){%
  \!ifnextchar({\!morxyl{#1}{#2}}{\!morObjB{#1}{#2}}}
\def\!morxyl#1#2(#3,#4){%
  \!ifnextchar[{\!mora{#1}{#2}{#3}{#4}}{\!mora{#1}{#2}{#3}{#4}[\number\distanciaobjmor,\number\distanciaobjmor]}}%
\def\!morObjA#1{%
 \let\pilhaaux=\pilha%
 \def\objPartida{#1}%
 \loop%
    \!topo\pilha%
    \!retira\pilha%
    \!compara{\id}{\objPartida}%
    \ifcompara \nvaziafalse \else \!pilhanvazia\pilha \fi%
   \ifnvazia%
 \repeat%
 \ifvazia%
  \ifdiferentes%
   Error: Incorrect label specification%
   \xaux=1%
   \yaux=1%
  \fi%
 \fi%
 \let\pilha=\pilhaaux%
 \!ifnextchar({\!morxyl{\number\xaux}{\number\yaux}}{\!morObjB{\number\xaux}{\number\yaux}}}
\def\!morObjB#1#2#3{%
  \x=#1
  \y=#2
 \def\objChegada{#3}%
 \let\pilhaaux=\pilha%
 \loop
    \!topo\pilha %
    \!retira\pilha%
    \!compara{\id}{\objChegada}%
    \ifcompara \nvaziafalse \else \!pilhanvazia\pilha \fi
   \ifnvazia
 \repeat
 \ifvazia
  \ifdiferentes%
   Error: Incorrect label specification
   \xaux=\x%
   \advance\xaux by \x%
   \yaux=\y%
   \advance\yaux by \y%
  \fi
 \fi
 \let\pilha=\pilhaaux
 \!ifnextchar[{\!mora{\number\x}{\number\y}{\number\xaux}{\number\yaux}}{\!mora{\number\x}{\number\y}{\number\xaux}{\number\yaux}[\number\distanciaobjmor,\number\distanciaobjmor]}}
\def\!mora#1#2#3#4[#5,#6]#7{%
  \!ifnextchar[{\!morb{#1}{#2}{#3}{#4}{#5}{#6}{#7}}{\!morb{#1}{#2}{#3}{#4}{#5}{#6}{#7}[1,\number\tipoarco] }}
\def\!morb#1#2#3#4#5#6#7[#8,#9]{\x=#1%
  \y=#2%
  \xl=#3%
  \yl=#4%
  \multiply \x by \expansao%
  \multiply \y by \expansao%
  \multiply \xl by \expansao%
  \multiply \yl by \expansao%
  \!quadrado{\number\x}{\number\xl}{\auxa}%
  \!quadrado{\number\y}{\number\yl}{\auxb}%
  \d=\auxa%
  \advance \d by \auxb%
  \!raiz{\d}{\d}%
  \auxa=#5
  \!compara{\objNulo}{\objPartida}%
  \ifdiferentes
   \!ajusta{\x}{\xl}{\y}{\yl}{\auxa}{\objPartida}%
   \ajustatrue
   \def\objPartida{}
  \fi
  \guardaauxa=\auxa
  \!ucoord{\number\x}{\number\xl}{\number\x}{\auxa}{\number\d}{\xa}{1}%
  \!ucoord{\number\y}{\number\yl}{\number\y}{\auxa}{\number\d}{\ya}{1}%
  \auxa=\d%
  \auxb=#6
  \!compara{\objNulo}{\objChegada}%
  \ifdiferentes
   \!ajusta{\x}{\xl}{\y}{\yl}{\auxb}{\objChegada}%
   \def\objChegada{}
  \fi
  \advance \auxa by -\auxb%
  \!ucoord{\number\x}{\number\xl}{\number\x}{\number\auxa}{\number\d}{\xb}{1}%
  \!ucoord{\number\y}{\number\yl}{\number\y}{\number\auxa}{\number\d}{\yb}{1}%
  \xmed=\xa%
  \advance \xmed by \xb%
  \divide \xmed by 2
  \ymed=\ya%
  \advance \ymed by \yb%
  \divide \ymed by 2
  \!distnomemor{\number\x}{\number\y}{\number\xl}{\number\yl}{#7}{\dnm}%
  \!ucoord{\number\y}{\number\yl}{\number\xmed}{\number\dnm}{\number\d}{\xc}{-#8}%
  \!ucoord{\number\x}{\number\xl}{\number\ymed}{\number\dnm}{\number\d}{\yc}{#8}%
\ifcase #9  
  \arrow <4pt> [.2,1.1] from {\xa} {\ya} to {\xb} {\yb}
\or  
  \setdashes
  \arrow <4pt> [.2,1.1] from {\xa} {\ya} to {\xb} {\yb}
  \setsolid
\or  
  \setlinear
  \plot {\xa} {\ya}  {\xb} {\yb} /
\or  
  \auxa=\guardaauxa
  \advance \auxa by 3%
 \!ucoord{\number\x}{\number\xl}{\number\x}{\number\auxa}{\number\d}{\xa}{1}%
 \!ucoord{\number\y}{\number\yl}{\number\y}{\number\auxa}{\number\d}{\ya}{1}%
 \!ucoord{\number\y}{\number\yl}{\number\xa}{3}{\number\d}{\xd}{-1}%
 \!ucoord{\number\x}{\number\xl}{\number\ya}{3}{\number\d}{\yd}{1}%
  \arrow <4pt> [.2,1.1] from {\xa} {\ya} to {\xb} {\yb}
  \circulararc -180 degrees from {\xa} {\ya} center at {\xd} {\yd}
\or  
  \auxa=3
 \!ucoord{\number\y}{\number\yl}{\number\xa}{\number\auxa}{\number\d}{\xmed}{-1}%
 \!ucoord{\number\x}{\number\xl}{\number\ya}{\number\auxa}{\number\d}{\ymed}{1}%
 \!ucoord{\number\y}{\number\yl}{\number\xa}{\number\auxa}{\number\d}{\xd}{1}%
 \!ucoord{\number\x}{\number\xl}{\number\ya}{\number\auxa}{\number\d}{\yd}{-1}%
  \arrow <4pt> [.2,1.1] from {\xa} {\ya} to {\xb} {\yb}
  \setlinear
  \plot {\xmed} {\ymed}  {\xd} {\yd} /
\fi
\auxa=\xl
\advance \auxa by -\x%
\ifnum \auxa=0 
  \put {#7} at {\xc} {\yc}
\else
  \auxb=\yl
  \advance \auxb by -\y%
  \ifnum \auxb=0 \put {#7} at {\xc} {\yc}
  \else 
    \ifnum \auxa > 0 
      \ifnum \auxb > 0
        \ifnum #8=1
          \put {#7} [rb] at {\xc} {\yc}
        \else 
          \put {#7} [lt] at {\xc} {\yc}
        \fi
      \else
        \ifnum #8=1
          \put {#7} [lb] at {\xc} {\yc}
        \else 
          \put {#7} [rt] at {\xc} {\yc}
        \fi
      \fi
    \else
      \ifnum \auxb > 0 
        \ifnum #8=1
          \put {#7} [rt] at {\xc} {\yc}
        \else 
          \put {#7} [lb] at {\xc} {\yc}
        \fi
      \else
        \ifnum #8=1
          \put {#7} [lt] at {\xc} {\yc}
        \else 
          \put {#7} [rb] at {\xc} {\yc}
        \fi
      \fi
    \fi
  \fi
\fi
}

\def\modifplot(#1{\!modifqcurve #1}
\def\!modifqcurve(#1,#2){\x=#1%
  \y=#2%
  \multiply \x by \expansao%
  \multiply \y by \expansao%
  \!start (\x,\y)
  \!modifQjoin}
\def\!modifQjoin(#1,#2)(#3,#4){\x=#1%
  \y=#2%
  \xl=#3%
  \yl=#4%
  \multiply \x by \expansao%
  \multiply \y by \expansao%
  \multiply \xl by \expansao%
  \multiply \yl by \expansao%
  \!qjoin (\x,\y) (\xl,\yl)             
  \!ifnextchar){\!fim}{\!modifQjoin}}
\def\!fim){\ignorespaces}

\def\setaxy(#1{\!pontosxy #1}
\def\!pontosxy(#1,#2){%
  \!maispontosxy}
\def\!maispontosxy(#1,#2)(#3,#4){%
  \!ifnextchar){\!fimxy#3,#4}{\!maispontosxy}}
\def\!fimxy#1,#2){\x=#1%
  \y=#2
  \multiply \x by \expansao
  \multiply \y by \expansao
  \xl=\x%
  \yl=\y%
  \aux=1%
  \multiply \aux by \auxa%
  \advance\xl by \aux%
  \aux=1%
  \multiply \aux by \auxb%
  \advance\yl by \aux%
  \arrow <4pt> [.2,1.1] from {\x} {\y} to {\xl} {\yl}}

\def\cmor#1 #2(#3,#4)#5{%
  \!ifnextchar[{\!cmora{#1}{#2}{#3}{#4}{#5}}{\!cmora{#1}{#2}{#3}{#4}{#5}[0] }}
\def\!cmora#1#2#3#4#5[#6]{%
  \ifcase #2
      \auxa=0
      \auxb=1
    \or
      \auxa=0
      \auxb=-1
    \or
      \auxa=1
      \auxb=0
    \or
      \auxa=-1
      \auxb=0
    \fi  
  \ifcase #6  
    \modifplot#1
    \setaxy#1
  \or  
    \setdashes
    \modifplot#1
    \setaxy#1
    \setsolid
  \or  
    \modifplot#1
  \fi  
  \x=#3%
  \y=#4%
  \multiply \x by \expansao%
  \multiply \y by \expansao%
  \put {#5} at {\x} {\y}}

\def\obj(#1,#2){%
  \!ifnextchar[{\!obja{#1}{#2}}{\!obja{#1}{#2}[Nulo]}}
\def\!obja#1#2[#3]#4{%
  \!ifnextchar[{\!objb{#1}{#2}{#3}{#4}}{\!objb{#1}{#2}{#3}{#4}[1]}}
\def\!objb#1#2#3#4[#5]{%
  \x=#1%
  \y=#2%
  \def\!pinta{\normalsize$\bullet$}
  \def\!nulo{Nulo}%
  \def\!arg{#3}%
  \!compara{\!arg}{\!nulo}%
  \ifcompara\def\!arg{#4}\fi%
  \multiply \x by \expansao%
  \multiply \y by \expansao%
  \setbox\caixa=\hbox{#4}%
  \!coloca{(\!arg)(#1,#2)(\number\ht\caixa,\number\wd\caixa,\number\dp\caixa)}{\pilha}%
  \auxa=\wd\caixa \divide \auxa by 131072 
  \advance \auxa by 5
  \auxb=\ht\caixa
  \advance \auxb by \number\dp\caixa
  \divide \auxb by 131072 
  \advance \auxb by 5
  \ifcase \tipografo    
    \put{#4} at {\x} {\y}
  \or                   
    \ifcase #5 
      \put{#4} at {\x} {\y}
    \or        
      \put{\!pinta} at {\x} {\y}
      \advance \y by \number\auxb  
      \put{#4} at {\x} {\y}
    \or        
      \put{\!pinta} at {\x} {\y}
      \advance \auxa by -2  
      \advance \auxb by -2  
      \advance \x by \number\auxa  
      \advance \y by \number\auxb  
      \put{#4} at {\x} {\y}   
    \or        
      \put{\!pinta} at {\x} {\y}
      \advance \x by \number\auxa  
      \put{#4} at {\x} {\y}   
    \or        
      \put{\!pinta} at {\x} {\y}
      \advance \auxa by -2  
      \advance \auxb by -2  
      \advance \x by \number\auxa  
      \advance \y by -\number\auxb  
      \put{#4} at {\x} {\y}   
    \or        
      \put{\!pinta} at {\x} {\y}
      \advance \y by -\number\auxb  
      \put{#4} at {\x} {\y}   
    \or        
      \put{\!pinta} at {\x} {\y}
      \advance \auxa by -2  
      \advance \auxb by -2  
      \advance \x by -\number\auxa  
      \advance \y by -\number\auxb  
      \put{#4} at {\x} {\y}   
    \or        
      \put{\!pinta} at {\x} {\y}
      \advance \x by -\number\auxa  
      \put{#4} at {\x} {\y}   
    \or        
      \put{\!pinta} at {\x} {\y}
      \advance \auxa by -2  
      \advance \auxb by -2  
      \advance \x by -\number\auxa  
      \advance \y by \number\auxb  
      \put{#4} at {\x} {\y}   
    \fi
  \or                   
    \ifcase #5 
      \put{#4} at {\x} {\y}
    \or        
      \put{\!pinta} at {\x} {\y}
      \advance \y by \number\auxb  
      \put{#4} at {\x} {\y}
    \or        
      \put{\!pinta} at {\x} {\y}
      \advance \auxa by -2  
      \advance \auxb by -2  
      \advance \x by \number\auxa  
      \advance \y by \number\auxb  
      \put{#4} at {\x} {\y}   
    \or        
      \put{\!pinta} at {\x} {\y}
      \advance \x by \number\auxa  
      \put{#4} at {\x} {\y}   
    \or        
      \put{\!pinta} at {\x} {\y}
      \advance \auxa by -2  
      \advance \auxb by -2
      \advance \x by \number\auxa  
      \advance \y by -\number\auxb 
      \put{#4} at {\x} {\y}   
    \or        
      \put{\!pinta} at {\x} {\y}
      \advance \y by -\number\auxb 
      \put{#4} at {\x} {\y}   
    \or        
      \put{\!pinta} at {\x} {\y}
      \advance \auxa by -2  
      \advance \auxb by -2
      \advance \x by -\number\auxa 
      \advance \y by -\number\auxb 
      \put{#4} at {\x} {\y}   
    \or        
      \put{\!pinta} at {\x} {\y}
      \advance \x by -\number\auxa 
      \put{#4} at {\x} {\y}   
    \or        
      \put{\!pinta} at {\x} {\y}
      \advance \auxa by -2  
      \advance \auxb by -2
      \advance \x by -\number\auxa 
      \advance \y by \number\auxb  
      \put{#4} at {\x} {\y}   
    \fi
   \else 
     \ifnum\auxa<\auxb 
       \aux=\auxb
     \else
       \aux=\auxa
     \fi
     \ifdim\wd\caixa<1em
       \dimen99 = 1em
       \aux=\dimen99 \divide \aux by 131072 
       \advance \aux by 5
     \fi
     \advance\aux by -2 
     \multiply\aux by 2 %
     \ifnum\aux<30
       \put{\circle{\aux}} [Bl] at {\x} {\y}
     \else
       \multiply\auxa by 2
       \multiply\auxb by 2
       \put{\oval(\auxa,\auxb)} [Bl] at {\x} {\y}
     \fi
     \put{#4} at {\x} {\y}
   \fi   
}

\catcode`!=12 

 \input miniltx
  \def\Gin@driver{pdftex.def}
  \input color.sty
  \input graphicx.sty
  \resetatcatcode

%
%

%
%
%
%

\def\Serif{cmr}
\def\SerifBold{cmbx}
\def\SerifItalics{cmti}
\def\SerifSlanted{cmsl}
\def\SerifBoldItalics{cmbxti}
\def\SansSerif{cmss}
\def\SansSerifBold{cmssbx}
\def\SansSerifItalics{cmssi}
\def\SansSerifSlanted{cmssi}
\def\Math{cmmi}
\def\Symbols{cmsy}
\def\MathBold{cmmib}
\def\MoreSymbols{cmex}
\def\Typewriter{cmtt}
\def\Gothic{eufm}
\def\Double{msbm}
\def\Relazioni{msam}

= 			\Serif10 			at 5pt
= 		\SerifBold10 		at 5pt
= 	\SerifItalics10 	at 5pt
=		\SerifSlanted10 	at 5pt
=	\SerifBoldItalics10	at 5pt
= 		\SansSerif10 		at 5pt
=	\SansSerifBold10	at 5pt
=	\SansSerifItalics10	at 5pt
=	\SansSerifSlanted10	at 5pt
=				\Math10				at 5pt
=			\MathBold10			at 5pt
=			\Symbols10			at 5pt
=		\MoreSymbols10		at 5pt
=		\Typewriter10		at 5pt
=			\Gothic10			at 5pt
=			\Double10			at 5pt

= 			\Serif10 			at 7pt
= 		\SerifBold10 		at 7pt
= 	\SerifItalics10 	at 7pt
=	\SerifSlanted10 	at 7pt
=\SerifBoldItalics10	at 7pt
= 		\SansSerif10 		at 7pt
= 	\SansSerifBold10 	at 7pt
=\SansSerifItalics10	at 7pt
=\SansSerifSlanted10	at 7pt
=			\Math10				at 7pt
=		\MathBold10			at 7pt
=			\Symbols10			at 7pt
=		\MoreSymbols10		at 7pt
=		\Typewriter10		at 7pt
=			\Gothic10			at 7pt
=			\Double10			at 7pt

= 			\Serif10 			at 8pt
= 		\SerifBold10 		at 8pt
= 	\SerifItalics10 	at 8pt
=	\SerifSlanted10 	at 8pt
=\SerifBoldItalics10	at 8pt
= 		\SansSerif10 		at 8pt
= 	\SansSerifBold10 	at 8pt
=\SansSerifItalics10 at 8pt
=\SansSerifSlanted10 at 8pt
=			\Math10				at 8pt
=		\MathBold10			at 8pt
=			\Symbols10			at 8pt
=		\MoreSymbols10		at 8pt
=		\Typewriter10		at 8pt
=			\Gothic10			at 8pt
=			\Double10			at 8pt

= 			\Serif10 			at 10pt
= 		\SerifBold10 		at 10pt
= 		\SerifItalics10 	at 10pt
=		\SerifSlanted10 	at 10pt
=	\SerifBoldItalics10	at 10pt
= 		\SansSerif10 		at 10pt
= 	\SansSerifBold10 	at 10pt
= 	\SansSerifItalics10 at 10pt
= 	\SansSerifSlanted10 at 10pt
=				\Math10				at 10pt
=			\MathBold10			at 10pt
=			\Symbols10			at 10pt
=		\MoreSymbols10		at 10pt
=		\Typewriter10		at 10pt
=			\Gothic10			at 10pt
=			\Double10			at 10pt
=			\Relazioni10			at 10pt

= 				\Serif10 			at 12pt
= 			\SerifBold10 		at 12pt
= 		\SerifItalics10 	at 12pt
=		\SerifSlanted10 	at 12pt
=	\SerifBoldItalics10	at 12pt
= 			\SansSerif10 		at 12pt
= 		\SansSerifBold10 	at 12pt
= 	\SansSerifItalics10 at 12pt
= 	\SansSerifSlanted10 at 12pt
=				\Math10				at 12pt
=			\MathBold10			at 12pt
=			\Symbols10			at 12pt
=		\MoreSymbols10		at 12pt
=			\Typewriter10		at 12pt
=				\Gothic10			at 12pt
=				\Double10			at 12pt

= 			\Serif10 			at 14pt
= 		\SerifBold10 		at 14pt
= 	\SerifItalics10 	at 14pt
=		\SerifSlanted10 	at 14pt
=	\SerifBoldItalics10	at 14pt
= 		\SansSerif10 		at 14pt
= 	\SansSerifBold10 	at 14pt
= \SansSerifSlanted10 at 14pt
= \SansSerifItalics10 at 14pt
=				\Math10				at 14pt
=			\MathBold10			at 14pt
=			\Symbols10			at 14pt
=		\MoreSymbols10		at 14pt
=		\Typewriter10		at 14pt
=			\Gothic10			at 14pt
=			\Double10			at 14pt

\def\NormalStyle{\parindent=5pt\parskip=3pt\normalbaselineskip=14pt%
\def\nt{\tenSerif}%
\def\rm{\fam0\tenSerif}%
\textfont0=\tenSerif\scriptfont0=\sevenSerif\scriptscriptfont0=\fiveSerif
\textfont1=\tenMath\scriptfont1=\sevenMath\scriptscriptfont1=\fiveMath
\textfont2=\tenSymbols\scriptfont2=\sevenSymbols\scriptscriptfont2=\fiveSymbols
\textfont3=\tenMoreSymbols\scriptfont3=\sevenMoreSymbols\scriptscriptfont3=\fiveMoreSymbols
\textfont\itfam=\tenSerifItalics\def\it{\fam\itfam\tenSerifItalics}%
\textfont\slfam=\tenSerifSlanted\def\sl{\fam\slfam\tenSerifSlanted}%
\textfont\ttfam=\tenTypewriter\def\tt{\fam\ttfam\tenTypewriter}%
\textfont\bffam=\tenSerifBold%
\def\bf{\fam\bffam\tenSerifBold}\scriptfont\bffam=\sevenSerifBold\scriptscriptfont\bffam=\fiveSerifBold%
\def\cal{\tenSymbols}%
\def\greekbold{\tenMathBold}%
\def\gothic{\tenGothic}%
\def\Bbb{\tenDouble}%
\def\LieFont{\tenSerifItalics}%
\nt\normalbaselines\baselineskip=14pt%
}

\def\TitleStyle{\parindent=0pt\parskip=0pt\normalbaselineskip=15pt%
\def\nt{\fourteenSansSerifBold}%
\def\rm{\fam0\fourteenSansSerifBold}%
\textfont0=\fourteenSansSerifBold\scriptfont0=\tenSansSerifBold\scriptscriptfont0=\eightSansSerifBold
\textfont1=\fourteenMath\scriptfont1=\tenMath\scriptscriptfont1=\eightMath
\textfont2=\fourteenSymbols\scriptfont2=\tenSymbols\scriptscriptfont2=\eightSymbols
\textfont3=\fourteenMoreSymbols\scriptfont3=\tenMoreSymbols\scriptscriptfont3=\eightMoreSymbols
\textfont\itfam=\fourteenSansSerifItalics\def\it{\fam\itfam\fourteenSansSerifItalics}%
\textfont\slfam=\fourteenSansSerifSlanted\def\sl{\fam\slfam\fourteenSerifSansSlanted}%
\textfont\ttfam=\fourteenTypewriter\def\tt{\fam\ttfam\fourteenTypewriter}%
\textfont\bffam=\fourteenSansSerif%
\def\bf{\fam\bffam\fourteenSansSerif}\scriptfont\bffam=\tenSansSerif\scriptscriptfont\bffam=\eightSansSerif%
\def\cal{\fourteenSymbols}%
\def\greekbold{\fourteenMathBold}%
\def\gothic{\fourteenGothic}%
\def\Bbb{\fourteenDouble}%
\def\LieFont{\fourteenSerifItalics}%
\nt\normalbaselines\baselineskip=15pt%
}

\def\PartStyle{\parindent=0pt\parskip=0pt\normalbaselineskip=15pt%
\def\nt{\fourteenSansSerifBold}%
\def\rm{\fam0\fourteenSansSerifBold}%
\textfont0=\fourteenSansSerifBold\scriptfont0=\tenSansSerifBold\scriptscriptfont0=\eightSansSerifBold
\textfont1=\fourteenMath\scriptfont1=\tenMath\scriptscriptfont1=\eightMath
\textfont2=\fourteenSymbols\scriptfont2=\tenSymbols\scriptscriptfont2=\eightSymbols
\textfont3=\fourteenMoreSymbols\scriptfont3=\tenMoreSymbols\scriptscriptfont3=\eightMoreSymbols
\textfont\itfam=\fourteenSansSerifItalics\def\it{\fam\itfam\fourteenSansSerifItalics}%
\textfont\slfam=\fourteenSansSerifSlanted\def\sl{\fam\slfam\fourteenSerifSansSlanted}%
\textfont\ttfam=\fourteenTypewriter\def\tt{\fam\ttfam\fourteenTypewriter}%
\textfont\bffam=\fourteenSansSerif%
\def\bf{\fam\bffam\fourteenSansSerif}\scriptfont\bffam=\tenSansSerif\scriptscriptfont\bffam=\eightSansSerif%
\def\cal{\fourteenSymbols}%
\def\greekbold{\fourteenMathBold}%
\def\gothic{\fourteenGothic}%
\def\Bbb{\fourteenDouble}%
\def\LieFont{\fourteenSerifItalics}%
\nt\normalbaselines\baselineskip=15pt%
}

\def\ChapterStyle{\parindent=0pt\parskip=0pt\normalbaselineskip=15pt%
\def\nt{\fourteenSansSerifBold}%
\def\rm{\fam0\fourteenSansSerifBold}%
\textfont0=\fourteenSansSerifBold\scriptfont0=\tenSansSerifBold\scriptscriptfont0=\eightSansSerifBold
\textfont1=\fourteenMath\scriptfont1=\tenMath\scriptscriptfont1=\eightMath
\textfont2=\fourteenSymbols\scriptfont2=\tenSymbols\scriptscriptfont2=\eightSymbols
\textfont3=\fourteenMoreSymbols\scriptfont3=\tenMoreSymbols\scriptscriptfont3=\eightMoreSymbols
\textfont\itfam=\fourteenSansSerifItalics\def\it{\fam\itfam\fourteenSansSerifItalics}%
\textfont\slfam=\fourteenSansSerifSlanted\def\sl{\fam\slfam\fourteenSerifSansSlanted}%
\textfont\ttfam=\fourteenTypewriter\def\tt{\fam\ttfam\fourteenTypewriter}%
\textfont\bffam=\fourteenSansSerif%
\def\bf{\fam\bffam\fourteenSansSerif}\scriptfont\bffam=\tenSansSerif\scriptscriptfont\bffam=\eightSansSerif%
\def\cal{\fourteenSymbols}%
\def\greekbold{\fourteenMathBold}%
\def\gothic{\fourteenGothic}%
\def\Bbb{\fourteenDouble}%
\def\LieFont{\fourteenSerifItalics}%
\nt\normalbaselines\baselineskip=15pt%
}

\def\SectionStyle{\parindent=0pt\parskip=0pt\normalbaselineskip=13pt%
\def\nt{\twelveSansSerifBold}%
\def\rm{\fam0\twelveSansSerifBold}%
\textfont0=\twelveSansSerifBold\scriptfont0=\eightSansSerifBold\scriptscriptfont0=\eightSansSerifBold
\textfont1=\twelveMath\scriptfont1=\eightMath\scriptscriptfont1=\eightMath
\textfont2=\twelveSymbols\scriptfont2=\eightSymbols\scriptscriptfont2=\eightSymbols
\textfont3=\twelveMoreSymbols\scriptfont3=\eightMoreSymbols\scriptscriptfont3=\eightMoreSymbols
\textfont\itfam=\twelveSansSerifItalics\def\it{\fam\itfam\twelveSansSerifItalics}%
\textfont\slfam=\twelveSansSerifSlanted\def\sl{\fam\slfam\twelveSerifSansSlanted}%
\textfont\ttfam=\twelveTypewriter\def\tt{\fam\ttfam\twelveTypewriter}%
\textfont\bffam=\twelveSansSerif%
\def\bf{\fam\bffam\twelveSansSerif}\scriptfont\bffam=\eightSansSerif\scriptscriptfont\bffam=\eightSansSerif%
\def\cal{\twelveSymbols}%
\def\bg{\twelveMathBold}%
\def\gothic{\twelveGothic}%
\def\Bbb{\twelveDouble}%
\def\LieFont{\twelveSerifItalics}%
\nt\normalbaselines\baselineskip=13pt%
}

\def\SubSectionStyle{\parindent=0pt\parskip=0pt\normalbaselineskip=13pt%
\def\nt{\twelveSansSerifItalics}%
\def\rm{\fam0\twelveSansSerifItalics}%
\textfont0=\twelveSansSerifItalics\scriptfont0=\eightSansSerifItalics\scriptscriptfont0=\eightSansSerifItalics%
\textfont1=\twelveMath\scriptfont1=\eightMath\scriptscriptfont1=\eightMath%
\textfont2=\twelveSymbols\scriptfont2=\eightSymbols\scriptscriptfont2=\eightSymbols%
\textfont3=\twelveMoreSymbols\scriptfont3=\eightMoreSymbols\scriptscriptfont3=\eightMoreSymbols%
\textfont\itfam=\twelveSansSerif\def\it{\fam\itfam\twelveSansSerif}%
\textfont\slfam=\twelveSansSerifSlanted\def\sl{\fam\slfam\twelveSerifSansSlanted}%
\textfont\ttfam=\twelveTypewriter\def\tt{\fam\ttfam\twelveTypewriter}%
\textfont\bffam=\twelveSansSerifBold%
\def\bf{\fam\bffam\twelveSansSerifBold}\scriptfont\bffam=\eightSansSerifBold\scriptscriptfont\bffam=\eightSansSerifBold%
\def\cal{\twelveSymbols}%
\def\greekbold{\twelveMathBold}%
\def\gothic{\twelveGothic}%
\def\Bbb{\twelveDouble}%
\def\LieFont{\twelveSerifItalics}%
\nt\normalbaselines\baselineskip=13pt%
}

\def\AuthorStyle{\parindent=0pt\parskip=0pt\normalbaselineskip=14pt%
\def\nt{\tenSerif}%
\def\rm{\fam0\tenSerif}%
\textfont0=\tenSerif\scriptfont0=\sevenSerif\scriptscriptfont0=\fiveSerif
\textfont1=\tenMath\scriptfont1=\sevenMath\scriptscriptfont1=\fiveMath
\textfont2=\tenSymbols\scriptfont2=\sevenSymbols\scriptscriptfont2=\fiveSymbols
\textfont3=\tenMoreSymbols\scriptfont3=\sevenMoreSymbols\scriptscriptfont3=\fiveMoreSymbols
\textfont\itfam=\tenSerifItalics\def\it{\fam\itfam\tenSerifItalics}%
\textfont\slfam=\tenSerifSlanted\def\sl{\fam\slfam\tenSerifSlanted}%
\textfont\ttfam=\tenTypewriter\def\tt{\fam\ttfam\tenTypewriter}%
\textfont\bffam=\tenSerifBold%
\def\bf{\fam\bffam\tenSerifBold}\scriptfont\bffam=\sevenSerifBold\scriptscriptfont\bffam=\fiveSerifBold%
\def\cal{\tenSymbols}%
\def\greekbold{\tenMathBold}%
\def\gothic{\tenGothic}%
\def\Bbb{\tenDouble}%
\def\LieFont{\tenSerifItalics}%
\nt\normalbaselines\baselineskip=14pt%
}

\def\AddressStyle{\parindent=0pt\parskip=0pt\normalbaselineskip=14pt%
\def\nt{\eightSerif}%
\def\rm{\fam0\eightSerif}%
\textfont0=\eightSerif\scriptfont0=\sevenSerif\scriptscriptfont0=\fiveSerif
\textfont1=\eightMath\scriptfont1=\sevenMath\scriptscriptfont1=\fiveMath
\textfont2=\eightSymbols\scriptfont2=\sevenSymbols\scriptscriptfont2=\fiveSymbols
\textfont3=\eightMoreSymbols\scriptfont3=\sevenMoreSymbols\scriptscriptfont3=\fiveMoreSymbols
\textfont\itfam=\eightSerifItalics\def\it{\fam\itfam\eightSerifItalics}%
\textfont\slfam=\eightSerifSlanted\def\sl{\fam\slfam\eightSerifSlanted}%
\textfont\ttfam=\eightTypewriter\def\tt{\fam\ttfam\eightTypewriter}%
\textfont\bffam=\eightSerifBold%
\def\bf{\fam\bffam\eightSerifBold}\scriptfont\bffam=\sevenSerifBold\scriptscriptfont\bffam=\fiveSerifBold%
\def\cal{\eightSymbols}%
\def\greekbold{\eightMathBold}%
\def\gothic{\eightGothic}%
\def\Bbb{\eightDouble}%
\def\LieFont{\eightSerifItalics}%
\nt\normalbaselines\baselineskip=14pt%
}

\def\AbstractStyle{\parindent=0pt\parskip=0pt\normalbaselineskip=12pt%
\def\nt{\eightSerif}%
\def\rm{\fam0\eightSerif}%
\textfont0=\eightSerif\scriptfont0=\sevenSerif\scriptscriptfont0=\fiveSerif
\textfont1=\eightMath\scriptfont1=\sevenMath\scriptscriptfont1=\fiveMath
\textfont2=\eightSymbols\scriptfont2=\sevenSymbols\scriptscriptfont2=\fiveSymbols
\textfont3=\eightMoreSymbols\scriptfont3=\sevenMoreSymbols\scriptscriptfont3=\fiveMoreSymbols
\textfont\itfam=\eightSerifItalics\def\it{\fam\itfam\eightSerifItalics}%
\textfont\slfam=\eightSerifSlanted\def\sl{\fam\slfam\eightSerifSlanted}%
\textfont\ttfam=\eightTypewriter\def\tt{\fam\ttfam\eightTypewriter}%
\textfont\bffam=\eightSerifBold%
\def\bf{\fam\bffam\eightSerifBold}\scriptfont\bffam=\sevenSerifBold\scriptscriptfont\bffam=\fiveSerifBold%
\def\cal{\eightSymbols}%
\def\greekbold{\eightMathBold}%
\def\gothic{\eightGothic}%
\def\Bbb{\eightDouble}%
\def\LieFont{\eightSerifItalics}%
\nt\normalbaselines\baselineskip=12pt%
}

\def\RefsStyle{\parindent=0pt\parskip=0pt%
\def\nt{\eightSerif}%
\def\rm{\fam0\eightSerif}%
\textfont0=\eightSerif\scriptfont0=\sevenSerif\scriptscriptfont0=\fiveSerif
\textfont1=\eightMath\scriptfont1=\sevenMath\scriptscriptfont1=\fiveMath
\textfont2=\eightSymbols\scriptfont2=\sevenSymbols\scriptscriptfont2=\fiveSymbols
\textfont3=\eightMoreSymbols\scriptfont3=\sevenMoreSymbols\scriptscriptfont3=\fiveMoreSymbols
\textfont\itfam=\eightSerifItalics\def\it{\fam\itfam\eightSerifItalics}%
\textfont\slfam=\eightSerifSlanted\def\sl{\fam\slfam\eightSerifSlanted}%
\textfont\ttfam=\eightTypewriter\def\tt{\fam\ttfam\eightTypewriter}%
\textfont\bffam=\eightSerifBold%
\def\bf{\fam\bffam\eightSerifBold}\scriptfont\bffam=\sevenSerifBold\scriptscriptfont\bffam=\fiveSerifBold%
\def\cal{\eightSymbols}%
\def\greekbold{\eightMathBold}%
\def\gothic{\eightGothic}%
\def\Bbb{\eightDouble}%
\def\LieFont{\eightSerifItalics}%
\nt\normalbaselines\baselineskip=10pt%
}

\def\ClaimStyle{\parindent=5pt\parskip=3pt\normalbaselineskip=14pt%
\def\nt{\tenSerifSlanted}%
\def\rm{\fam0\tenSerifSlanted}%
\textfont0=\tenSerifSlanted\scriptfont0=\sevenSerifSlanted\scriptscriptfont0=\fiveSerifSlanted
\textfont1=\tenMath\scriptfont1=\sevenMath\scriptscriptfont1=\fiveMath
\textfont2=\tenSymbols\scriptfont2=\sevenSymbols\scriptscriptfont2=\fiveSymbols
\textfont3=\tenMoreSymbols\scriptfont3=\sevenMoreSymbols\scriptscriptfont3=\fiveMoreSymbols
\textfont\itfam=\tenSerifItalics\def\it{\fam\itfam\tenSerifItalics}%
\textfont\slfam=\tenSerif\def\sl{\fam\slfam\tenSerif}%
\textfont\ttfam=\tenTypewriter\def\tt{\fam\ttfam\tenTypewriter}%
\textfont\bffam=\tenSerifBold%
\def\bf{\fam\bffam\tenSerifBold}\scriptfont\bffam=\sevenSerifBold\scriptscriptfont\bffam=\fiveSerifBold%
\def\cal{\tenSymbols}%
\def\greekbold{\tenMathBold}%
\def\gothic{\tenGothic}%
\def\Bbb{\tenDouble}%
\def\LieFont{\tenSerifItalics}%
\nt\normalbaselines\baselineskip=14pt%
}


%
%


\def\ModeYes{yes}
\def\ModeNo{no}

\def\ModeUndef{undefined}


\def\nx{\noexpand}
\def\ni{\noindent}
\def\newpage{\vfill\eject}

\def\ss{\vskip 5pt}
\def\ms{\vskip 10pt}
\def\bs{\vskip 20pt}

 \def\,{\mskip\thinmuskip}
 \def\!{\mskip-\thinmuskip}
 \def\>{\mskip\medmuskip}
 \def\;{\mskip\thickmuskip}

%
% ... 
%

\def\refsModePost{post}
\def\refsModeAuto{auto}

\def\dbRefsSatusModeOk{ok}
\def\dbRefsSatusModeError{error}
\def\dbRefsSatusModeWarning{warning}


\newcount\BNUM
\BNUM=0

\def\refs{}

\def\SetModePost{\xdef\refsMode{\refsModePost}}			
\SetModePost

\def\dbRefsStatusOk{%
	\xdef\dbRefsStatus{\dbRefsSatusModeOk}%
	\xdef\dbRefsError{\ModeNo}%
	\xdef\dbRefsWarning{\ModeNo}%
	\xdef\dbRefsInfo{\ModeNo}%
}

\def\dbRefs{%
}

\def\dbRefsGet#1{%
	\xdef\found{N}\xdef\ikey{#1}\dbRefsStatusOk%
	\xdef\key{\ModeUndef}\xdef\tag{\ModeUndef}\xdef\tail{\ModeUndef}%
	\dbRefs%
}

\def\NextRefsTag{%
	\global\advance\BNUM by 1%
}
\def\ShowTag#1{{\bf [#1]}}

\def\dbRefsInsert#1#2{%
\dbRefsGet{#1}%
\if\found Y %
   \xdef\dbRefsStatus{\dbRefsSatusModeWarning}%
   \xdef\dbRefsWarning{record is already there}%
   \xdef\dbRefsInfo{record not inserted}%
\else%
   \toks2=\expandafter{\dbRefs}%
   \ifx\refsMode\refsModeAuto \NextRefsTag
    \xdef\dbRefs{%
   	\the\toks2 \nx\xdef\nx\dbx{#1}%
	\nx\ifx\nx\ikey %
		\nx\dbx\nx\xdef\nx\found{Y}%
		\nx\xdef\nx\key{#1}%
		\nx\xdef\nx\tag{\the\BNUM}%
		\nx\xdef\nx\tail{#2}%
	\nx\fi}%
	\global\xdef\refs{\refs \ss\ni[\the\BNUM]\ #2\par}
   \fi%
   \ifx\refsMode\refsModePost 
    \xdef\dbRefs{%
   	\the\toks2 \nx\xdef\nx\dbx{#1}%
	\nx\ifx\nx\ikey %
		\nx\dbx\nx\xdef\nx\found{Y}%
		\nx\xdef\nx\key{#1}%
		\nx\xdef\nx\tag{\ModeUndef}%
		\nx\xdef\nx\tail{#2}%
	\nx\fi}%
   \fi%
\fi%
}

\def\dbRefsEdit#1#2#3{\dbRefsGet{#1}%
\if\found N 
   \xdef\dbRefsStatus{\dbRefsSatusModeError}%
   \xdef\dbRefsError{record is not there}%
   \xdef\dbRefsInfo{record not edited}%
\else%
   \toks2=\expandafter{\dbRefs}%
   \xdef\dbRefs{\the\toks2%
   \nx\xdef\nx\dbx{#1}%
   \nx\ifx\nx\ikey\nx\dbx %
	\nx\xdef\nx\found{Y}%
	\nx\xdef\nx\key{#1}%
	\nx\xdef\nx\tag{#2}%
	\nx\xdef\nx\tail{#3}%
   \nx\fi}%
\fi%
}

\def\bib#1#2{\RefsStyle\dbRefsInsert{#1}{#2}%
	\ifx\dbRefsStatus\dbRefsSatusModeWarning %
		\message{^^J}%
		\message{WARNING: Reference [#1] is doubled.^^J}%
	\fi%
}

\def\ref#1{\dbRefsGet{#1}%
\ifx\found N %
  \message{^^J}%
  \message{ERROR: Reference [#1] unknown.^^J}%
  \ShowTag{??}%
\else%
	\ifx\tag\ModeUndef \NextRefsTag%
		\dbRefsEdit{#1}{\the\BNUM}{\tail}%
		\dbRefsGet{#1}%
		\global\xdef\refs{\refs \ss\ni [\tag]\ \tail\par}
	\fi
	\ShowTag{\tag}%
\fi%
}

\def\ShowBiblio{\ms\Ensure{\SectionEnsure}%
{\SectionStyle\ni References}%
{\RefsStyle\refs}%
}

\newcount\CHANGES
\CHANGES=0
\def\AuxFile{7}
\def\PreventDoubleOn{\xdef\PreventDoubleLabel{\ModeYes}}

\PreventDoubleOn

\def\StoreLabel#1#2{\xdef\itag{#2}
 \ifx\PreModeStatus\ModeNo %
   \message{^^J}%
   \errmessage{You can't use Check without starting with OpenPreMode (and finishing with ClosePreMode)^^J}%
 \else%
   \immediate\write\AuxFile{\nx\dbLabelPreInsert{#1}{\itag}}%
   \dbLabelGet{#1}%
   \ifx\itag\tag %
   \else%
	\global\advance\CHANGES by 1%
 	\xdef\itag{(?.??)}%
    \fi%
   \fi%
}

\def\PreModeStatus{\ModeNo}

\def\ClosePreMode{\immediate\closeout\AuxFile%
  \ifnum\CHANGES=0%
	\message{^^J}%
	\message{**********************************^^J}%
	\message{**  NO CHANGES TO THE AuxFile  **^^J}%
	\message{**********************************^^J}%
 \else%
	\message{^^J}%
	\message{**************************************************^^J}%
	\message{**  PLAEASE TYPESET IT AGAIN (\the\CHANGES)  **^^J}%
    \errmessage{**************************************************^^ J}%
  \fi%
  \edef\PreModeStatus{\ModeNo}%
}

\def\dbLabelSatusModeOk{ok}

\def\dbLabelSatusModeWarning{warning}

\def\dbLabelStatusOk{%
	\xdef\dbLabelStatus{\dbLabelSatusModeOk}%
	\xdef\dbLabelError{\ModeNo}%
	\xdef\dbLabelWarning{\ModeNo}%
	\xdef\dbLabelInfo{\ModeNo}%
}

\def\dbLabel{%
}

\def\dbLabelGet#1{%
	\xdef\found{N}\xdef\ikey{#1}\dbLabelStatusOk%
	\xdef\key{\ModeUndef}\xdef\tag{\ModeUndef}\xdef\pre{\ModeUndef}%
	\dbLabel%
}

\def\ShowLabel#1{%
 \dbLabelGet{#1}%
 \ifx\tag \ModeUndef %
 	\global\advance\CHANGES by 1%
 	(?.??)%
 \else%
 	\tag%
 \fi%
}

\def\dbLabelPreInsert#1#2{\dbLabelGet{#1}%
\if\found Y %
  \xdef\dbLabelStatus{\dbLabelSatusModeWarning}%
   \xdef\dbLabelWarning{Label is already there}%
   \xdef\dbLabelInfo{Label not inserted}%
   \message{^^J}%
   \errmessage{Double pre definition of label [#1]^^J}%
\else%
   \toks2=\expandafter{\dbLabel}%
    \xdef\dbLabel{%
   	\the\toks2 \nx\xdef\nx\dbx{#1}%
	\nx\ifx\nx\ikey %
		\nx\dbx\nx\xdef\nx\found{Y}%
		\nx\xdef\nx\key{#1}%
		\nx\xdef\nx\tag{#2}%
		\nx\xdef\nx\pre{\ModeYes}%
	\nx\fi}%
\fi%
}

\def\dbLabelInsert#1#2{\dbLabelGet{#1}%
\xdef\itag{#2}%
\dbLabelGet{#1}%
\if\found Y %
	\ifx\tag\itag %
	\else%
	   \ifx\PreventDoubleLabel\ModeYes %
		\message{^^J}%
		\errmessage{Double definition of label [#1]^^J}%
	   \else%
		\message{^^J}%
		\message{Double definition of label [#1]^^J}%
	   \fi%
	\fi%
   \xdef\dbLabelStatus{\dbLabelSatusModeWarning}%
   \xdef\dbLabelWarning{Label is already there}%
   \xdef\dbLabelInfo{Label not inserted}%
\else%
   \toks2=\expandafter{\dbLabel}%
    \xdef\dbLabel{%
   	\the\toks2 \nx\xdef\nx\dbx{#1}%
	\nx\ifx\nx\ikey %
		\nx\dbx\nx\xdef\nx\found{Y}%
		\nx\xdef\nx\key{#1}%
		\nx\xdef\nx\tag{#2}%
		\nx\xdef\nx\pre{\ModeNo}%
	\nx\fi}%
\fi%
}


\newcount\PART
\newcount\CHAPTER
\newcount\SECTION
\newcount\SUBSECTION
\newcount\FNUMBER

\PART=0
\CHAPTER=0
\SECTION=0
\SUBSECTION=0	
\FNUMBER=0

\def\LastPart{\ModeUndef}
\def\LastChapter{\ModeUndef}
\def\LastSection{\ModeUndef}
\def\LastSubSection{\ModeUndef}
\def\LastClaim{\ModeUndef}
\def\Last{\ModeUndef}

\newdimen\TOBOTTOM
\newdimen\LIMIT

\def\Ensure#1{\ \par\ \immediate\LIMIT=#1\immediate\TOBOTTOM=\the\pagegoal\advance\TOBOTTOM by -\pagetotal%
\ifdim\TOBOTTOM<\LIMIT\newpage \else%
\vskip-\parskip\vskip-\parskip\vskip-\baselineskip\fi}

\def\PartLabel{\the\PART}
\def\NewPart#1{\global\advance\PART by 1%
         \bs\ni{\PartStyle  Part \PartLabel:}
         \bs\ni{\PartStyle #1}\newpage%
         \CHAPTER=0\SECTION=0\SUBSECTION=0\FNUMBER=0%
         \gdef\Left{#1}%
         \global\edef\Last{\PartLabel}%
         \global\edef\LastPart{\PartLabel}%
         \global\edef\LastChapter{\ModeUndef}%
         \global\edef\LastSection{\ModeUndef}%
         \global\edef\LastSubSection{\ModeUndef}%
         \global\edef\LastClaim{\ModeUndef}}
\def\ChapterLabel{\the\CHAPTER}
\def\NewChapter#1{\global\advance\CHAPTER by 1%
         \bs\ni{\ChapterStyle  Chapter \ChapterLabel: #1}\ms%
         \SECTION=0\SUBSECTION=0\FNUMBER=0%
         \gdef\Left{#1}%
         \global\edef\Last{\ChapterLabel}%
         \global\edef\LastChapter{\ChapterLabel}%
         \global\edef\LastSection{\ModeUndef}%
         \global\edef\LastSubSection{\ModeUndef}%
         \global\edef\LastClaim{\ModeUndef}}
\def\SectionEnsure{3cm}
\def\NewSection#1{\Ensure{\SectionEnsure}\gdef\SectionLabel{\the\SECTION}\global\advance\SECTION by 1%
         \ms\ni{\SectionStyle  \SectionLabel.\ #1}\ss%
         \SUBSECTION=0\FNUMBER=0%
         \gdef\Left{#1}%
         \global\edef\Last{\SectionLabel}%
         \global\edef\LastSection{\SectionLabel}%
         \global\edef\LastSubSection{\ModeUndef}%
         \global\edef\LastClaim{\ModeUndef}}
\def\NewAppendix#1#2{\Ensure{\SectionEnsure}\gdef\SectionLabel{#1}\global\advance\SECTION by 1%
         \bs\ni{\SectionStyle  Appendix \SectionLabel.\ #2}\ss%
         \SUBSECTION=0\FNUMBER=0%
         \gdef\Left{#2}%
         \global\edef\Last{\SectionLabel}%
         \global\edef\LastSection{\SectionLabel}%
         \global\edef\LastSubSection{\ModeUndef}%
         \global\edef\LastClaim{\ModeUndef}}
\def\Acknowledgements{\Ensure{\SectionEnsure}\gdef\SectionLabel{}%
         \ms\ni{\SectionStyle  Acknowledgments}\ss%
         \SECTION=0\SUBSECTION=0\FNUMBER=0%
         \gdef\Left{}%
         \global\edef\Last{\ModeUndef}%
         \global\edef\LastSection{\ModeUndef}%
         \global\edef\LastSubSection{\ModeUndef}%
         \global\edef\LastClaim{\ModeUndef}}
\def\SubSectionEnsure{2cm}
\def\SubSectionLabel{\ifnum\SECTION>0 \the\SECTION.\fi\the\SUBSECTION}
\def\NewSubSection#1{\Ensure{\SubSectionEnsure}\global\advance\SUBSECTION by 1%
         \ms\ni{\SubSectionStyle #1}\ss%
         \global\edef\Last{\SubSectionLabel}%
         \global\edef\LastSubSection{\SubSectionLabel}}
\def\SetNumberingModeN{\def\ClaimLabel{(\the\FNUMBER)}}
\def\SetNumberingModeSN{\def\ClaimLabel{(\ifnum\SECTION>0 \SectionLabel.\fi%
      \the\FNUMBER)}}
\def\SetNumberingModeCSN{\def\ClaimLabel{(\ifnum\CHAPTER>0 \the\CHAPTER.\fi%
      \ifnum\SECTION>0 \SectionLabel.\fi%
      \the\FNUMBER)}}

\def\NewClaim{\global\advance\FNUMBER by 1%
    \ClaimLabel%
    \global\edef\LastClaim{\ClaimLabel}%
    \global\edef\Last{\ClaimLabel}}

\def\HideLabels{\xdef\ShowLabelsMode{\ModeNo}}
\HideLabels

\def\fn{\eqno{\NewClaim}} 
\def\fl#1{%
\ifx\ShowLabelsMode\ModeYes%
 \eqno{{\buildrel{\hbox{\AbstractStyle[#1]}}\over{\hfill\NewClaim}}}%
\else%
 \eqno{\NewClaim}%
\fi%
\dbLabelInsert{#1}{\ClaimLabel}}
\def\fprel#1{\global\advance\FNUMBER by 1\StoreLabel{#1}{\ClaimLabel}%
\ifx\ShowLabelsMode\ModeYes%
\eqno{{\buildrel{\hbox{\AbstractStyle[#1]}}\over{\hfill.\itag}}}%
\else%
 \eqno{\itag}%
\fi%
}

\def\cl#1{\global\advance\FNUMBER by 1\dbLabelInsert{#1}{\ClaimLabel}%
\ifx\ShowLabelsMode\ModeYes%
${\buildrel{\hbox{\AbstractStyle[#1]}}\over{\hfill\ClaimLabel}}$%
\else%
  $\ClaimLabel$%
\fi%
}
\def\cprel#1{\global\advance\FNUMBER by 1\StoreLabel{#1}{\ClaimLabel}%
\ifx\ShowLabelsMode\ModeYes%
${\buildrel{\hbox{\AbstractStyle[#1]}}\over{\hfill.\itag}}$%
\else%
  $\itag$%
\fi%
}

\def\Note{\ms\leftskip 3cm\rightskip 1.5cm\AbstractStyle}
\def\endNote{\par\leftskip 2cm\rightskip 0cm\NormalStyle\ss}


\parindent=7pt
\leftskip=2cm
\newcount\SideIndent
\newcount\SideIndentTemp
\SideIndent=0
\newdimen\SectionIndent
\SectionIndent=-8pt

\def\sidebar{\vrule height15pt width.2pt }
\def\endcorner{\hbox{\hbox{\vrule height6pt width.2pt}\vbox to6pt{\vfill\hbox
to4pt{\leaders\hrule height0.2pt\hfill}}}}
\def\begincorner{\hbox{\hbox{\vrule height6pt width.2pt}\vbox to6pt{\hbox
to4pt{\leaders\hrule height0.2pt\hfill}}}}
\def\endbegincorner{\hbox{\vbox to15pt{\endcorner\vskip-6pt\begincorner\vfill}}}
\def\SideShow{\SideIndentTemp=\SideIndent \ifnum \SideIndentTemp>0 
\loop\sidebar\hskip 2pt \advance\SideIndentTemp by-1\ifnum \SideIndentTemp>1 \repeat\fi}

\def\BeginSection{{\vbadness 100000 \par\ni\hskip\SectionIndent%
\SideShow\vbox to 15pt{\vfill\begincorner}}\global\advance\SideIndent by1\vskip-10pt}

\def\EndSection{{\vbadness 100000 \par\ni\global\advance\SideIndent by-1%
\hskip\SectionIndent\SideShow\vbox to15pt{\endcorner\vfill}\vskip-10pt}}

\def\EndBeginSection{{\vbadness 100000\par\ni%
\global\advance\SideIndent by-1\hskip\SectionIndent\SideShow
\vbox to15pt{\vfill\endbegincorner}}%
\global\advance\SideIndent by1\vskip-10pt}

\def\ShowBeginCorners#1{%
\SideIndentTemp =#1 \advance\SideIndentTemp by-1%
\ifnum \SideIndentTemp>0 %
\vskip-15truept\hbox{\kern 2truept\vbox{\hbox{\begincorner}%
\ShowBeginCorners{\SideIndentTemp}\vskip-3truept}}%
\fi%
}

\def\ShowEndCorners#1{%
\SideIndentTemp =#1 \advance\SideIndentTemp by-1%
\ifnum \SideIndentTemp>0 %
\vskip-15truept\hbox{\kern 2truept\vbox{\hbox{\endcorner}%
\ShowEndCorners{\SideIndentTemp}\vskip 2truept}}%
\fi%
}

\def\BeginSections#1{{\vbadness 100000 \par\ni\hskip\SectionIndent%
\SideShow\vbox to 15pt{\vfill\ShowBeginCorners{#1}}}\global\advance\SideIndent by#1\vskip-10pt}

\def\EndSections#1{{\vbadness 100000 \par\ni\global\advance\SideIndent by-#1%
\hskip\SectionIndent\SideShow\vbox to15pt{\vskip15pt\ShowEndCorners{#1}\vfill}\vskip-10pt}}

\def\EndBeginSections#1#2{{\vbadness 100000\par\ni%
\global\advance\SideIndent by-#1%
\hbox{\hskip\SectionIndent\SideShow\kern-2pt%
\vbox to15pt{\vskip15pt\ShowEndCorners{#1}\vskip4pt\ShowBeginCorners{#2}}}}%
\global\advance\SideIndent by#2\vskip-10pt}




%
%


\def\al{\alpha}
\def\be{\beta}
\def\de{\delta}
\def\ga{\gamma}

\def\ep{\epsilon}

\def\om{\omega}
\def\si{\sigma}

\def\Ga{\Gamma}

\def\Om{\Omega}
\def\Si{\Sigma}


 \def\calC{{\hbox{\cal C}}}

 \def\calS{{\hbox{\cal S}}}




 \def\R{{\hbox{\Bbb R}}}

 \def\E{{\hbox{\Bbb E}}}

 \def\R{{\hbox{\Bbb R}}}


\def\Lor{{\hbox{Lor}}}

\def\id{{\hbox{\rm id}}}
\def\diag{{\hbox{diag}}}

\def\curl{{\hbox{curl}}}

\def\ip{\hbox to4pt{\leaders\hrule height0.3pt\hfill}\vbox to8pt{\leaders\vrule width0.3pt\vfill}\kern 2pt}
 
\def\del{\partial}
\def\na{\nabla}

\def\arr{\rightarrow}

\def\then{\Rightarrow}

%
%

\def\THEOREM{\ClaimStyle\ni{\bf Theorem: }}

\def\ENDTHEOREM{\NormalStyle}

\def\cases#1{\left\{\eqalign{#1}\right.}
\NormalStyle
\SetNumberingModeSN
\PreventDoubleOn

\long\def\title#1{\centerline{\TitleStyle\ni#1}}

\long\def\author#1{\ms\centerline{\AuthorStyle by {\it #1}}}

\long\def\address#1{\ss\centerline{\AddressStyle #1}\par}
\long\def\moreaddress#1{\centerline{\AddressStyle #1}\par}
\def\abstract{\ms\leftskip 3cm\rightskip .5cm\AbstractStyle{\bf \ni Abstract:}\ }
\def\endabstract{\par\leftskip 2cm\rightskip 0cm\NormalStyle\ss}

\SetNumberingModeSN

\def\frac[#1/#2]{\hbox{$#1\over#2$}}
\def\Frac[#1/#2]{{#1\over#2}}
\def\({\left(}
\def\){\right)}
\def\[{\left[}
\def\]{\right]}
\def\^#1{{}^{#1}_{\>\cdot}}
\def\_#1{{}_{#1}^{\>\cdot}}
\def\Label=#1{{\buildrel {\hbox{\fiveSerif \ShowLabel{#1}}}\over =}}
\def\<{\kern -1pt}


\def\ExpandAllCNotes{\long\def\CNote##1{%
\BeginSection
	\Note%
 		##1%
	\endNote%
\EndSection%
}}
\ExpandAllCNotes
%
%
%
%


\def\blue#1{\textcolor{blue}{#1}}

\def\frame#1{\vbox{\hrule\hbox{\vrule\vbox{\kern2pt\hbox{\kern2pt#1\kern2pt}\kern2pt}\vrule}\hrule\kern-4pt}}

\def\Box to #1#2#3{\frame{\vtop{\hbox to #1{\hfill #2 \hfill}\hbox to #1{\hfill #3 \hfill}}}}


\bib{BeigMurchadha}{R.Being, N.\'o Murchadha,
{\it The Poincar\'e Group as the Symmetry Group of Canonical General Relativity},
Annals of Physics {\bf 174}, 1987, 463-498 
}

\bib{RovelliBook}{C. Rovelli, {\it Quantum Gravity}, Cambridge University Press, Cambridge, 2004
}

\bib{Simon1}{L.Fatibene, S.Garruto, {\it The Cauchy problem in general relativity: an algebraic characterization}, Classical and Quantum Gravity, 32, 23, p. 235010}

\bib{Taylor}{M. Taylor, {\it Partial Differential Equations III}, Springer, New York (1996).
}

\bib{Hebey1}{E. Hebey, {\it Sobolev spaces on Riemannian manifolds}, Lecture Notes in Mathematics, vol. 1635, Springer-Verlag, Berlin, 1996.
}

\bib{Hebey2}{E. Hebey, {\it Nonlinear analysis on manifolds: Sobolev spaces and inequalities}, Courant Lect. Notes
Math., Vol. 5, Courant Institute of Mathematical Sciences, New York University, New York, 1999.
}

\bib{Christodoulou}{D.Christodoulou, 
{\it On the global initial value problem and the issue of singularities}, 
Class. Quantum Gravity 16 (1999).
}

\bib{Book}{L. Fatibene, M. Francaviglia, {\it Natural and gauge natural formalism for classical field theories. A geometric perspective including spinors and gauge theories}, Kluwer Academic Publishers, Dordrecht, (2003).
}

\bib{Gourgoulhon}{E. Gourgoulhon. 
{\it 3+1 Formalism and Bases of Numerical Relativity}, arXiv:gr-qc/0703035, (2007).
}

\bib{CB1}{
Y.Choquet-Bruhat, 
{\it Th\'eor\`eme d'existence pour certains syst\`emes d'\'equations aux d\'eriv\'ees partielles non lin\' eaires},
 Acta Math. 88(1952).
}

\bib{CB2}{
Y.Choquet-Bruhat, R.Geroch, 
{\it Global Aspects of the Cauchy Problem in General Relativity}, 
Comm. Math. Phys. 14 (1969).
}

\bib{CB3}{
Y.Choquet-Bruhat, D.Christodoulou, M.Francaviglia, 
{\it Probl\`eme de Cauchy sur une vari\'et\'e}, 
C.R. Acad Sci. Paris S\'er. A-B 287(5) (1978).
}

\bib{Moncrief}{
V.Moncrief, D.M.Eardley, The global existence problem and cosmic censorship in general relativity, Gen. Rel. Grav. 13 (1981).
}

\bib{York}{
J.W. York, 
{\it Gravitational Degrees of Freedom and the Initial-Value Problem}, Phys. Rev. Lett. 26 (1971).
}

\bib{ADM}{R. Arnowitt, S. Deser and C. W. Misner, in: {\it Gravitation: An Introduction to Current Research}, L. Witten ed. Wyley, 227, (New York, 1962); gr-qc/0405109
}

\bib{NostroADM}{L. Fatibene, M. Ferraris, M. Francaviglia, L.Lusanna, {\it ADM Pseudotensors, Conserved Quantities and Covariant Conservation Laws in General Relativity}, arXiv:gr-qc/1007.4071.
}

\bib{HE}{S.W. Hawking, G.F.R. Ellis {\it The Large Scale Structure of Space-Time}, Cambridge:
Camblidge University Press, 1975.
}



\def\ubal{\underline{\al}\kern1pt}
\def\obal{\overline{\al}\kern1pt}

\def\ubR{\underline{R}\kern1pt}
\def\obR{\overline{R}\kern1pt}
\def\ubom{\underline{\om}\kern1pt}
\def\obxi{\overline{\xi}\kern1pt}
\def\ubu{\underline{u}\kern1pt}
\def\ube{\underline{e}\kern1pt}
\def\obe{\overline{e}\kern1pt}

\def\New#1{{\blue{#1}}}

\bib{Gotay1}{M.J. Gotay, J. Isenberg, J.E. Marsden, R.Montgomery, {\it Momentum Maps and Classical Relativistic Fields, Part I: Covariant Field Theory}, arXiv:physics/9801019v2.}

\bib{Gotay2}{M.J. Gotay, J. Isenberg, J.E. Marsden, {\it Momentum Maps and Classical Relativistic Fields, Part II: Canonical Analysis of Field Theories}, arXiv:math-ph/0411032.}

\bib{Gotay3}{M.J. Gotay, J.E. Marsden, {\it Momentum Maps and Classical Relativistic Fields, Part III: Gauge Symmetries and Initial Value
Constraints}.}

\bib{Gotay4}{J.E. Marsden, S. Pekarsky, S. Shkoller, M. West, , {\it Variational methods, multisymplectic geometry and continuum mechanics}, Journal of Geometry and Physics, Volume 38, Issues 3-4, (2001).}

\bib{Sardanash}{G. Sardanashvily,
{\it Generalized Hamiltonian Formalism for Field Theory},
World Scienti?c, Singapore (1995)
}

\bib{Kijo}{J Kijowski, 
{\it A finite dimensional canonical formalism in the classical field theory}, 
Comm Math Phys 30 (1973) 99-128}

\bib{FF}{M Ferraris, M Francaviglia, 
{\it The Lagrangian approach to conserved quantities in General relativity}, 
in: Mechanics, Analysis and Geometry: 200 Years after Lagrange, pf 451-488, ed M Francaviglia (Elsevier Sci Publ, Amsterdam: 1991.)}

\bib{Rov}{C.Rovelli,
{\it Covariant hamiltonian formalism for field theory: Hamilton-Jacobi equation on the space $G$},
Lect. Notes Phys.633:36-62 (2003); arXiv:gr-qc/0207043
}

\bib{Gibbons}{G. W. Gibbons,
{\it The time symmetric initial value problem for black holes},
Comm. in Math. Phys.  {\bf 27}(2)  (1972), pp 87-102}

\bib{Ringstrom}{H. Ringstrom, 
{\it The Cauchy Problem in General Relativity}, 
European Mathematical Society, Zurich, 2009}

\bib{I1}{
J. Isenberg, J. Nester, 
{\it The Effect of Gravitational Interaction on Classical Fields: A Hamilton Dirac Analysis}, 
Ann. Phys. 107, 56-81, 1977.}

\bib{I2}{
J. Isenberg, J. Nester, 
{\it Canonical Gravity, Invited review article in General Relativity and Gravitation}, 
in: Einstein Centenary Volume, edited by A. Held) Plenum, New York, 1980}

\bib{I3}{J.Isenberg,
{\it Initial Value Problem in General Relativity},
in: The Springer Handbook of Spacetime, A.Ashtekar, V.Petkov, (Eds.) 
Springer-Verlag Berlin Heidelberg (2014); arXiv:1304.1960 [gr-qc]
}

\NormalStyle

\title{Principal Symbol of Euler-Lagrange Operators}

\author{L.Fatibene$^{a,b}$, S.Garruto$^{a,b}$}

\address{$^a$ Department of Mathematics, University of Torino (Italy)}
\moreaddress{$^b$ INFN - Sezione Torino}

\abstract  
We shall introduce the principal symbol for \New{quite a general class of (quasi linear)} Euler-Lagrange operators and use them to characterise well-posed initial value problems \New{in gauge  covariant field theories}.
We shall clarify how  constraints can arise in covariant Lagrangian  theories by extending the standard treatment in GR
\New{and without resorting to Hamiltonian formalism}.   
Finally as an example of application, we sketch a quantisation procedure based on what is done in LQG \New{by framing it is a more general context which applies to general gauge covariant field theories}.
\endabstract

\NewSection{Introduction}

Since most physically relevant field theories are degenerate due to gauge symmetries,
the standard approach to canonical quantization goes through a constraint analysis of their Hamiltonian formulation; see for example \ref{RovelliBook} for application to Loop Quantum Gravity (LQG).

Unfortunately, there is no consensus on a covariant Hamiltonian framework for field theories and different approaches often differ on details; see \ref{Gotay1}, \ref{Gotay2}, \ref{Gotay3}, \ref{Gotay4}, \ref{Kijo}, \ref{FF}, \ref{Sardanash}, \ref{Rov}.

Moreover, while some are deduced directly from Hamilton equations, other constraints emerge from the Poisson structure which is non-canonical on the Lagrangian side. 
Accordingly, it is not clear whether it is always possible and where to search for constraints in Lagrangian formalism \New{without resorting to some equivalent symplectic structure};
see \ref{I1}, \ref{I2}.

The same constraint analysis used for quantization is however used for the analysis of well-posedness of Cauchy problems; see \ref{CB1}, \ref{CB2}, \ref{CB3}, \ref{I3}, \ref{Ringstrom}, \ref{Christodoulou}, \ref{Gibbons}. 
Since Hamilton equations are equivalent to Euler-Lagrange equations it is almost trivial to notice that the same information should be available in the Lagrangian framework.

In \ref{Simon1} we reviewed the analysis of Einstein equations and their initial value problem.
The analysis relies on a formulation of the Cauchy theorem which claims that a Cauchy problem is well-posed if the principal symbol of the PDE operator is {\it symmetric hyperbolic}; see \ref{Simon1} or \ref{Taylor} and below.
This allows a complete algebraic analysis based on the structure of principal symbol.
In the case of standard GR, the principal symbol of (the evolution part of) Einstein equations turns out not to be symmetric.
That is where gauge fixing enters the game. 
By choosing harmonic coordinates (see \ref{CB1}, \ref{CB2}, \ref{CB3},   \ref{Christodoulou}, \ref{Gibbons}, 
\ref{Simon1}, \ref{Gourgoulhon})  one can obtain a symmetric hyperbolic problem for which Cauchy theorem holds true (see \ref{Taylor} and references quoted therein, as well as \ref{HE}) and the initial problem is well-posed.

However, one should argue that since the principal symbol is invariant and intrinsically defined changing coordinates should not affect its symmetry properties, as in fact we shall show below in general, i.e.~without even assuming fields are sections of a vector bundle.
In view of invariance of principal symbols there is no transformation on the configuration bundle (and hence no coordinate transformation on the base) which can make symmetric a principal  symbol which was not originally symmetric.

In order to solve this apparent contradiction (and better understanding the role of harmonic gauge in GR) we shall show that one can split the (covariant) equation in two (non-covariant) parts, one providing a symmetric hyperbolic (hence well-posed) problem, the other produced by 
the antisymmetric part either becomes a further constraint or it is satisfied in a particular class of coordinate systems.  
Solving the non-covariant problem in one of such coordinate systems provides solutions of the original covariant problem and then by changing coordinates a solution of the original problem is found in any coordinate system.
 
Here we have to check that this mechanism works in general for a Lagrangian field theory. The general analysis of course forgets about details and it is probably easier to follow than the particular analysis of standard GR. On the other hand, standard GR can be considered as an example of application
of the general framework here introduced. 

Although most of what follows can be easily extended to general differential operators between bundles we shall specialise to operators which come from a variational principle (see \ref{Book}). This of course includes  the application to natural theories and GR and it turns out to show that the general analytical approach (see \ref{Taylor}) does not in fact depend on a metric structure on fields or a linear structure \New{as it is often assumed in the literature}.

As a matter of fact this framework provides a way of analysing Dirac-Bergman constraints on a purely Lagrangian basis.

In Section 2 we shall fix notation.
In Section 3 we shall introduce the principal symbol and review its globality. 
\New{We have to stress that our result here is obtained without introducing a linear structure on fields which are assumed to be sections of a general bundle, not a linear bundle.
Morover, the duality between fields and equations naturally defines symmetry without resorting to an inner product between fields.}
In Section 4 we shall review ADM formalism \New{framing it in bundle theory. We also compute transformation laws of the coefficients of the principal symbol of first and second order quasi-linear operators.}
In Section 5 we propose two schemes for associating to the original operator a symmetric hyperbolic one, possibly introducing new constraints or partially fixing coordinate gauge.
\New{This procedure is quite general and it applies to all field theories of interest to the fundamental physics}.
In Section 6 we briefly describe few examples.
In Section 7 we briefly state how this structure can be adapted to a quantisation scheme proposed by Rovelli \New{for standard GR}; see \ref{RovelliBook}.
\New{Our aim here is just to show that the same procedure proposed for LQG does in fact apply in general. 
Of course, we are not claiming that one can trivially quantise any gauge covariant theory (for example implementing constraints as quantum operators is definitely not trivial and needs to be done on a case by case basis).
We believe however that extending the procedure to more general gauge covariant theories would better clarify to what extent LQG is peculiar of standard GR.}

The whole framework presented here is just a sketch and a lot of work and further investigation is needed in order to better understand its application scope (which here is supported by few but relevant standard cases only). We briefly discuss perspectives in Section 7.

\NewSection{Notation}

We shall hereafter deal with Lagrangian field theories. Fields are (global) sections of a bundle $\calC=(C, M, \pi, F)$, called {\it configuration bundle}.
In general we shall not assume the configuration bundle to be a vector (or affine) bundle.
In GR the configuration bundle $\Lor(M)$ is not a vector bundle (though it is a sub-bundle of a vector bundle $\Lor(M)\subset S_2(M)$, namely the bundle of symmetric tensors of rank $2$).
The base manifold $M$ is thought as a (connected,  oriented,  paracompact) spacetime manifold of dimension $\dim(M)=m$.

We shall assume fibered coordinates $(x^\mu, y^I)$ on $C$ and general transition functions in the form
$$
\cases{
&x'^\mu= x'^\mu(x)\cr
&y'^I= Y^I(x, y )\cr
}
\fl{TFC}$$

In order to deal with PDE one has to introduce jet bundles $J^k\calC$. 
Any fibered coordinates on $C$ induces a {\it natural fibered ``coordinate'' system} $j^k y= (x^\mu, y^I, y^I_\mu,\dots , y^I_{\mu_1\dots \mu_k})$
on the jet prolongations. The coordinates $(y^I_\mu, y^I_{\mu_1\mu_2}, \dots , y^I_{\mu_1\dots \mu_k})$ are meant to represent partial derivatives of fields $y^I(x)$ with respect to independent variables $x^\mu$ and they are accordingly considered as symmetric in lower indices.

Transition functions \ShowLabel{TFC} on the configuration  bundle  $\calC$ 
are prolonged to the jet bundles. In particular,  one has
$$
y'^I_{\mu_1\dots \mu_{k}}= J^I_J(x, y) y^J_{\si_1\dots\si_k} \bar J^{\si_1}_{\mu_1}(x)\dots  \bar J^{\si_{2k}}_{\mu_{k}}(x)
+ q^I_{\mu_1\dots \mu_{k}}(j^{k-1} y)
\fl{TFJ}$$
where $j^{k-1} y= (x^\mu, y^I, y^I_\mu, \dots , y^I_{\mu_1\dots \mu_{k-1}})$ are natural fibered coordinates on $J^{k-1}\calC$.
Here we set  $J^I_J= \del_J Y^I(x, y)$ and $J^I_\mu= \del_\mu Y^I(x, y)$ for the Jacobians of the fiber transformation and $J^{\si}_{\mu}(x)= \del_\mu x'^\si(x)$ for the Jacobian along the base coordinates. The bar over Jacobians denotes inverse matrices.

Let us remark that, for a general bundle $C$, transformations \ShowLabel{TFJ} are affine transformations, meaning that the jet bundle $\pi^k_{k-1}: J^k C\arr J^{k-1}C$ is always affine
even when the bundle $C$ is not affine or linear.
\New{It is precisely because of this (rather trivial) remark that one can define quasi-linear equations without assuming a linear structure on fields.}

\NewSection{Field equations in a Lagrangian field theory}

In general, differential operators are maps between a suitable jet prolongation bundle $J^{d}\calC$ into some fixed target vector bundle $E$. 
The corresponding differential equation is identified with the kernel of the map considered as  a submanifold of the jet prolongation, namely, $S\subset J^d\calC$.

\New{Let us remark that by regarding differential operators as (global) bundle morphisms and differential equations as submanifolds 
one sets up a correspondence between geometrical properties and the analytical properties which are independent of the coordinates and fields used for local expressions.
For example, it would be convenient to have a notion of well-posedness which were independent of coordinates just because it is easier to check if it is verified.
}

In variational calculus differential operators are not generic but obtained by Euler-Lagrange morphisms. 
They are hence in a specific form which is particularly well-suited to discuss principal symbols of differential operators. 
In particular they show that principal symbols and their relevant properties are canonically defined and independent of any metric defined on the base manifold (i.e.~the spacetime) or on the standard fiber (i.e.~the space of fields). Moreover, it is independent of any linear (or affine) structure defined on the configuration bundle.  

If $\calC$ is assumed as the configuration bundle of a Lagrangian field theory, then Euler-Lagrange operator of a $k$-order Lagrangian is a map
$$
\tilde \E:J^{2k}\calC \arr V^\ast(C)\otimes A_m(M) 
\fn$$
where $J^{2k}\calC$ is the jet prolongation of the configuration bundle $\calC$, $(V(C), C, p,  \R^k)$ is the vector bundle of vertical vectors on $\calC$, 
$V^\ast(C)$ denotes its dual, $A_m(M)$ is, by an abuse of notation, the pull-back of the bundle of $m$-form over $M$ along the projection $\pi:C\arr M$
(and hence should be denoted precisely as $\pi^\ast(A_m(M))$ to denote a bundle over $C$). 
The tensor product $V^\ast(C)\otimes A_m(M) $ is hence to be considered as a the tensor product of vector bundles over $C$.

\Note
\New{
If $(C, M, \pi, F)$ is a bundle over $M$ and $\phi:N\arr M$ is a manifold map, then one can define the {\it pull-back bundle} $(\phi^\ast C, N, \pi^\ast, F)$
which is a bundle over $N$ and it has the same standard fiber of the original bundle $C$.
The total space $\phi^\ast C$ is defined as
$$
\phi^\ast C=\{(z, c)\in N\times C:  \pi(c)=\phi(z)\}
\fn$$
and one can show it is automatically a bundle over $N$. 
If $(x^\mu, y^I)$ are fibered coordinates on $C$ then fiber coordinates on $\phi^\ast C$ are in the form $(z^i, y^I)$,
i.e.~the same fields over another space.
}

\New{
In the case of the bundle $A_m(M)$  of $m$-forms over a manifold $M$ one has coordinates $(x^\mu, \om)$ for the point $\Om= \om d\si\in A_m(M)$ (where $d\si$ is the basis for $m$-forms
induced by base coordinates).
Accordingly, the bundle $(\pi^\ast(A_m(M)), C, \pi^\ast, \R)$ has coordinates $(x^\mu, y^I, \om)$
and the bundle $\pi^\ast\(V^\ast(C)\otimes A_m(M)\)$ has coordinates $(x^\mu, y^I, e_I)$ for the objects as $e=e_I  \bar d y^I\otimes d\si\in V^\ast(C)\otimes A_m(M)$
where we set $\bar d y^I$ for the dual basis of vertical vectors $\del_I$.
}

\New{
Of course, pull-back bundles should not be confused with the pull-back of forms or tensorial objects.
}
\endNote

If the Lagrangian is degenerate (as it often happens in field theory) the Euler-Lagrange morphism can actually be of lower order, i.e.~it \New {happens to be the lift of a lower order  morphism}
$$
\E:J^{d}\calC \arr V^\ast(C)\otimes A_m(M) 
\fn$$
for some $d\le 2k$ \New{which makes the following diagram commutative
$$
\begindc{\commdiag}[1]
\obj(290,110)[C]{$V^\ast(C)\otimes A_m(M)$}
\obj(110,60)[JC]{$J^d\calC$}
\obj(110,110)[*JC]{$J^{2k}\calC$}
\mor{*JC}{C}{$\tilde \E$}
\mor{*JC}{JC}{$\pi^{2k}_d$}
\mor{JC}{C}{$\E$}
\enddc
\fn$$
so that $\tilde \E=\E\circ \pi^{2k}_d$.}

Since $\pi^{d}_{d-1}:J^{d}\calC\arr J^{d-1}\calC$ is an affine bundle, one can intrinsically define quasi-linear operators on any configuration bundle, 
i.e.~also  when $\calC$ is not assumed to be an affine or a vector bundle.

Non-degenerate variational equations are always quasi-linear in view of their variational origin. Also in degenerate cases all equations which are used in fundamental physics are quasi-linear although one can provide (somehow artificial) examples of Lagrangians with field equations which are not quasi-linear.

\Note
For example, if one considers the Lagrangian (on $(\R\times \R^2, \R^2, \pi, \R)$ with coordinates $(x^1, x^2, y)$)
$$
L=\Frac[y/(y_1)^2+ (y_2)^2](y_{11} y_{22}-(y_{12})^2)
\fn$$
field equations are obtained as
$$
\Frac[y_{11} y_{22}-(y_{12})^2/(y_1)^2+ (y_2)^2]=0
\fn$$
which are second order (instead of fourth-order) and not quasi-linear.
\endNote

In any event, we shall hereafter restrict to Lagrangians of order $k$ which produce quasi-linear Euler-Lagrange equations of order $d\le 2k$.
Accordingly, one has the Euler-Lagrange operator in the form
$$
\E= \E_I(j^{d} y)\> \bar d y^I\otimes d\si
\qquad
\E_I(j^{d}y)= e_{IJ}^{\al_1\dots \al_{d}} y^J_{\al_1\dots \al_{d}} + b_I(j^{d-1}y)
\fn$$

The coefficient of the leading term transforms as
$$
\New{e'{}_{IJ}^{\al_1\dots \al_{d}}= \bar J(x) \bar J_I^K(x, y) \bar J_J^H(x, y) e_{KH}^{\be_1\dots \be_{d}}  J^{\al_1}_{\be_1}(x)\dots   J^{\al_{d}}_{\be_{d}}(x)}
\fl{PrincipalTermTL}$$
where $J(x)$ denotes the determinant of the Jacobian $J_\mu^\al$.

\Note
Depending on the theory, one can have $e_{IJ}^{\al_1\dots \al_{d}}$ depending on $(x^\mu, y^I)$, or on $(j^{n}y)$ with $0\le n\le d-1$ and each of it is preserved by transition functions \ShowLabel{TFC}.
On the contrary, in general, coefficients  $e_{IJ}^{\al_1\dots \al_{d}}$ depending on $(x)$,  are not preserved by transition functions. If $e_{IJ}^{\al_1\dots \al_{d}}$ depend on $(x)$ in one trivialization, they might depend on $(x, y)$ in another trivialization.

In GR the configuration bundle $\Lor(M)$ is not a vector bundle.
The leading term $e_{IJ}^{\al \be}$ is a function of $(g^{\mu\nu})$ which is preserved by transition functions.
\endNote

Notice that $e_{KH}^{\be_1\dots \be_{d}}$ is symmetric with respect to $(KH)$ (non-degenerate, respectively)  if and only if $e'{}_{IJ}^{\al_1\dots \al_{d}}$ is also symmetric  with respect to $(IJ)$ (non-degenerate, respectively).
In other words, if the morphism is not symmetric  (non-degenerate, respectively)  in one trivialization it cannot be made symmetric  (non-degenerate, respectively) by changing trivialization.
In particular it cannot be made symmetric  (non-degenerate, respectively) by changing coordinates on $M$ even though $\calC$ is a natural bundle and changes of coordinates on $M$ induce functorially changes of trivializations.

Consider now a bundle morphism (over the identity $\id_C: C\arr C$)
$$
\si: S^{d}(T^\ast M)\arr V^\ast(\calC)\otimes V^\ast(\calC)\otimes  A_m(M)
\fl{PrincipalSymbolMorphism}$$
where $S^{d}(T^\ast M)$ is the symmetrized $d$-power of cotangent bundle (pulled-back on $C$ along the projection $\pi:C\arr M$).
\New{Depending on the case, if the leading coefficients depend on derivatives of fields then the pull-back of $S^{d}(T^\ast M)$  is understood to be on a suitable jet prolongation 
$J^n\calC$ along the projection $\pi^n:J^n\calC\arr M$. For example in standard GR the coefficients do not depend on derivatives and hence they are pulled back on $C$.}

The bundle $S^{d}(T^\ast M)$ (pulled back on $J^nC$ along $\pi^n:J^nC\arr M$, for a suitable $n$, possibly $n=0$) has fibered coordinates $(x^\mu,  y^I,\dots, y^I_{\al_1\dots \al_n}, \xi_{\al_1\dots \al_{d}})$ (which are meant to be symmetric in the lower indices 
$(\al_1\dots \al_{d})$). In view of transition functions the fibered morphism $\si$ is expressed as
$$
\si(j^ny, \xi_{\al_1\dots \al_{d}})=\si_{IJ}^{\al_1\dots \al_{d}}(j^n y)   \xi_{\al_1\dots \al_{d}} \bar d y^I \otimes \bar d y^J\otimes d\si
\fn$$
with $0\le n\le d-1$
and the coefficient $\si_{IJ}^{\al_1\dots \al_{d}}$ transforms as in \ShowLabel{PrincipalTermTL}.

Thus there is a one-to-one correspondence between leading terms of global differential operators $\E$ and global linear bundle morphisms $\si$.

\Note
If $\calC=(E, M, \pi, V)$ is an affine bundle then transition functions are in the form
$$
\cases{
&x'^\mu= x'^\mu(x)\cr
&v'^I= A^I_J(x) v^J + B^I(x)\cr
}
\fn$$
and the Jacobian $J^I_J$ specifies to $J^I_J(x, y)=  A^I_J(x) $ and it does depend on the base coordinates only.
Accordingly, in this case one can also consider differential operators with a leading term which depends on base coordinates only (namely, having $ e_{IJ}^{\al_1\dots \al_{d}}(x)$), being 
this property independent of the trivialization on vector bundles. 

For the principal symbol morphism $\si: S^{d}(T^\ast M)\arr V^\ast(\calC)\otimes V^\ast(\calC)\otimes  A_m(M)$  one can use the canonical isomorphism $V(\calC)\simeq E\oplus E$ and the projection onto the second factor to identify the bundles $V^\ast(\calC)\simeq E\oplus E^\ast$.  
\New{By further assuming an inner product on the fiber of the configuration bundle $E$ one has non-canonical isomorphisms $E\simeq E^\ast$ so that
$V^\ast(\calC)\otimes V^\ast(\calC)\otimes  A_m(M)\simeq E\oplus E^\ast\otimes E \otimes  A_m(M)$.}
Thus the principal symbol can be written in the form
$$
\si: S^{d}(T^\ast M)\otimes E\arr  E\otimes  A_m(M)
\fn$$
which is often found in the literature. We prefer the form \ShowLabel{PrincipalSymbolMorphism} which is more canonical being it available for any configuration bundle, not only on vector and affine bundles.
Moreover, the form \ShowLabel{PrincipalSymbolMorphism} makes it easier to discuss symmetry properties and non-degeneracy.

\endNote

The morphism $\si$ is called the {\it principal symbol morphism}.  If the image of $\si$ in contained into symmetric bilinear forms $S^2(V^\ast(\calC))\subset V^\ast(\calC)\otimes V^\ast(\calC)$ (a property which is independent of the trivialization, as noticed above) then the differential operator is called {\it symmetric}. If the image of $\si$ (except the zero section in $S^{2k}(T^\ast M)$)  is contained into non-degenerate symmetric bilinear forms  then the differential operator is called {\it elliptic}.

For any quasi-linear operator one can define the principal symbol and {\it characteristic directions}  are the covectors $\xi$
such that $\si(\xi\otimes\dots\otimes \xi)=0$. That describes the characteristic distribution.
One can show that field equations constrain possible values of solutions along characteristics directions. Accordingly, Cauchy hypersurfaces should be chosen so that no characteristic direction lies on them. 

\NewSection{ADM decomposition}

An {\it ADM splitting} of the spacetime $M$ is a bundle  $\calS=(M, \R, t, \Si)$ which foliates the spacetime $M$ in a $1$-parameter family of $3$-manifolds isomorphic to $\Si$. This is a general result since any foliation in hypersurfaces can be considered as a bundle (provided the leaves are all diffeomorphic); see \ref{ADM}, \ref{NostroADM}.

\Note
Notice that being the base $\R$ of the bundle $\calS$ contractible the bundle is trivial, then $M\simeq \Si\times \R$ and $M$ is called a {\it globally hyperbolic} spacetime.

The fibers $\Si_s=t^{-1}(s)\subset M$ are called {\it synchronous hypersurfaces} or {\it space hypersurfaces} at time $s$. The map $t:M\arr \R$ is usually thought as a map attaching the time at each event that happens and events on the same fiber happen at the same time and are hence synchronous.

Of course, different observers may define different notion of synchronization and then different ADM splittings.
\endNote

The spacetime $M$ can be covered with  coordinates $(t, x^a)$ fibered with respect to the ADM splitting. Such coordinate systems are called {\it ADM coordinates}
or {\it ADM observers}. The transition functions between ADM observers are in the form
$$
\cases{
&t'=t'(t)\cr
&x'^a= x'^a(t, x)\cr
}
\fn$$
and any (globally hyperbolic) spacetime can always be covered with a family of ADM observers which share the same definition of time.
ADM coordinates are not the most general coordinate systems on $M$ since in general one can consider observers which do not share the same ADM splitting.

\Note 
Notice that a connection of the ADM splitting (is a rank $1$, hence integrable distribution and it) defines integral curves which connect events on different leaves $\Si_s$ of the ADM foliation.
Such a congruence of curves establishes isomorphisms among spaces at different times and eventually defines a notion of {\it events at rest} with respect to the ADM spitting.
\endNote

Given a quasi-linear  differential operator $\E$  of order $d=1$  and an ADM splitting the operator can be written in ADM coordinates as
$$
e_{IJ} y^J_{0} +  e_{IJ}^{a}y^J_{a}  + b_I(t, x, y)=0
\fl{FOO}$$
where the coefficients $e_{IJ}(t, x, y)$ and $e_{IJ}^{a}(t, x, y)$ are meant to be independent of first derivatives of fields.

A first order operator is called {\it symmetric hyperbolic} if $e_{IJ}$ is \New{positive definite} (and hence invertible, the inverse being denoted by $e^{IJ}$) and if $e_{IJ}^{a}$ is symmetric in $(IJ)$.

An important result states that for first order symmetric hyperbolic operators the Cauchy theorem holds true, i.e.~there exists a unique solution for each initial condition, namely:
\ss

\THEOREM
Let $e_{IJ}$ be a non-degenerate \New{positive definite} bilinear form and let $e^a_{IJ}$ be symmetric in indices $(IJ)$ for all $a$, then
under these hypotheses, given the initial data in $H^k(M)$, with $k > \frac[m/2] + 1$, the existence and uniqueness is ensured {in an open interval $I\subset \R$ } and in a suitable Sobolev space, which depends on the regularity of the initial data (see \ref{Taylor}).
\ENDTHEOREM
\ss

Here $H^k(M)$ is the Sobolev space of $k$ order defined over $M$, see~\ref{Taylor}, \ref{Hebey1} and \ref{Hebey2}.
\New{This theorem does encapsulate all analytical results leaving us with a purely algebraic condition on the principal symbol.}

Let us investigate the transformation rules of principal symbol in order to determine whether being symmetric hyperbolic is a property of the operator or it depends on the ADM coordinate system.

\NewSubSection{First order operators}

Let us start with first order operators, i.e. operators in the form
$$
e_{IJ} y^J_{0} + e_{IJ}^{a}y^J_{a} + b_I(x, y)=0
\fn$$

Transition functions  with respect to changes of ADM coordinates read as
$$
\cases{
&y'^I_0= \bar J_0^0 J^I_J y^J_0  + \bar J_0^a J^I_J y^J_a+  \bar J_0^a J^I_a
\cr
&y'^I_a= \bar J_a^b (J^I_J y^J_b + J^I_b)  \qquad (\bar J^0_a=0\hbox{ for ADM coordinates})\cr
}\fn$$

Thus the operator in the new coordinates reads 
$$
J J^I_L \(e'_{IJ} y'^J_{0} + e'{}_{IJ}^{a}y'^J_{a} + b'_I( x, y)\) =0
\fn$$
and it relates to the operator in the old coordinates as
$$
J \bar J_0^0 J^I_L e'_{IJ}  J^J_K y^K_0 +J J^I_L \( e'_{IJ} \bar J_0^b + e'{}_{IJ}^{a}\bar J_a^b\) J^J_K  y^K_b + J J^I_L\( b'_I+ e'_{IJ} \bar J_0^\al J^J_\al+ e'{}_{IJ}^{a}\bar J_a^b J^J_b\)=0
\fn$$
Accordingly, the coefficients transform as
$$
\cases{
& e_{LK}= J \bar J_0^0 e'_{IJ} J^I_L  J^J_K\cr
&e_{LK}^b= J \(   e'_{IJ} \bar J_0^b + e'{}_{IJ}^{a}\bar J_a^b\) J^I_L J^J_K\cr
&b_L=  J J^I_L\( b'_I+ e'_{IJ} \bar J_0^\al J^J_\al+ e'{}_{IJ}^{a}\bar J_a^b J^J_b\)\cr
}
\fn$$

Accordingly, if the operator is symmetric hyperbolic in a system of ADM coordinates, it is symmetric hyperbolic in every ADM coordinate systems.
In fact, the antisymmetric part of the symbol $e_{[LK]}^b$ transforms as a tensor object
$$
e'{}_{[IJ]}^a=  \bar J J_b^a \>  e{}_{[LK]}^{b} \bar J^L_I  \bar J^K_J\
\fn$$

\NewSubSection{Second order operators}

For general quasi-linear second order operators, i.e.~operators in the form
$$
e_{IJ} y^J_{00} + e_{IJ}^{a}y^J_{0a} + e_{IJ}^{ab}y^J_{ab} + b_I(j^1y)=0
\fl{SOO}$$
a similar result holds true.

This operator is {\it symmetric hyperbolic} if $e_{IJ}$ and $e^{ab}_{IJ}$ are symmetric in the lower indices $(IJ)$, invertible and non-degenerate \New{positive definite} and
$e_{IJ}^{a}$ is symmetric in the lower indices $(IJ)$.
For second-order symmetric hyperbolic operators the Cauchy theorem holds true, i.e.~there exists a unique solution for each initial condition;
see the Appendix.

Transition functions of second derivatives  with respect to changes of ADM coordinates read as
$$
\cases{
&y'^I_{00}= \bar J_0^0 \bar J_0^0  J^I_J y^J_{00} 
+ 2\bar J_0^0 \bar J_0^a  J^I_J y^J_{0a}+ \bar J_0^b \bar J_0^a  J^I_J y^J_{ab}  + q'^I_{00}(j^1 y)   \cr
&y'^I_{0a}=  \bar J_a^c \bar J_0^d  J^I_J y^J_{d c}  + \bar J_a^c \bar J_0^0  J^I_J y^J_{0 c}  + q'^I_{0a}(j^1 y) \cr
&y'^I_{ab}=  \bar J_b^c \bar J_a^d  J^I_J y^J_{d c}  + q'^I_{ab}(j^1 y) \cr
}\fn$$

The operator in the new coordinates reads as
$$
J J^I_L \(e'_{IJ} y'^J_{00} + e'^{a}_{IJ}y'^J_{0a} + e'^{ab}_{IJ}y'^J_{ab} + b'_I(j^1 y')\) =0
\fn$$

and it relates to the operator in the old coordinates as
$$
\eqalign{
&J J^I_L  \(\bar J_0^0 \bar J_0^0 e'_{IJ}   J^J_K \) y^K_{00} 
 + J J^I_L  \(  2\bar J_0^0 \bar J_0^c  e'_{IJ} J^J_K+ \bar J_a^c \bar J_0^0 e'^{a}_{IJ} J^J_K \) y^K_{0 c}   +\cr
&+J J^I_L  \( \bar J_0^c \bar J_0^d  e'_{IJ}  J^J_K
+ \bar J_a^c \bar J_0^d e'^{a}_{IJ} J^J_K  
+ \bar J_a^c \bar J_b^d  e'^{ab}_{IJ}  J^J_K\) y^K_{cd}  +\cr
&+J J^I_L   \(b'_I(j^1 y') + e'_{IJ}  q'^J_{00}(j^1 y)  + e'^{a}_{IJ} q'^J_{0a}(j^1 y) +e'^{ab}_{IJ}   q'^J_{ab}(j^1 y)\)  =0
}
 \fn$$
Accordingly, the leading coefficients transform as
$$
\cases{
& e_{LK}   = J  \( \bar J_0^0 \bar J_0^0 e'_{IJ}\) J^I_L    J^J_K    \cr
& e_{LK}^c  = J  \( 2\bar J_0^0 \bar J_0^c  e'_{IJ} + \bar J_a^c \bar J_0^0 e'^{a}_{IJ} \)J^I_L  J^J_K\cr
&e_{LK}^{cd}  = J  \( \bar J_0^c \bar J_0^d  e'_{IJ}  
+ \bar J_a^c \bar J_0^d e'^{a}_{IJ} 
+ \bar J_a^c \bar J_b^d  e'^{ab}_{IJ}  \) J^I_L J^J_K \cr
}
\fn$$
Let us remark that $e'_{IJ}$ is \New{positive definite}  iff $e_{LK}$ is \New{positive definite}  in any other trivialization.
Analogously, as far as $e'_{IJ}$ is symmetric, $e_{LK}^c$ is symmetric in $(LK)$ iff $ e'^{a}_{IJ}$ is symmetric in $(IJ)$ in any other trivialization.
Finally,  as far as $e'_{IJ}$ and $e'^{a}_{IJ} $ are symmetric in $(IJ)$,
then  $e_{LK}^{cd}$ is symmetric in $(LK)$ iff $ e'^{ab}_{IJ}$ is symmetric in $(IJ)$ in any other trivialization.

Let us also stress that in standard GR one has that $e_{LK}$ is symmetric (and \New{positive definite}), $e_{LK}^c=0$ so that $e'^{a}_{IJ}$ is symmetric in any trivialization,
and hence $e'^{ab}_{IJ}$ cannot be made symmetric by changing trivialization.

Accordingly, at least when $e_{LK}$ and $e_{LK}^c$ are symmetric in $(LK)$,
if the operator is symmetric hyperbolic in a system of ADM coordinates, it is symmetric hyperbolic in every ADM coordinate systems.
In fact, the antisymmetric part of the symbol $e_{[LK]}^{cd}$ transforms as a tensor object
$$
e'{}_{[IJ]}^{ab}=  \bar J J_c^a J_d^b \>  e{}_{[LK]}^{cd} \bar J^L_I \bar J^K_J\
\fn$$

\NewSection{Symmetrisation of operators}

In the previous Section we showed that an Euler-Lagrange operator cannot be made symmetric hyperbolic by changing coordinates or trivialization if it is not symmetric hyperbolic in the beginning (except the specific cases in which $e^a_{IJ}$ is not symmetric, which however does not include standard GR). 
On the other hand, standard GR provides us with an example of a field theory with Euler-Lagrange operator which is not symmetric hyperbolic; see \ref{Simon1}.

However, there are at least two ways in which one can construct a new symmetric hyperbolic operator out of the original Euler-Lagrange operator.
Let us discuss these methods on a first order operator.

Given a quasi-linear first order operator 
$$
e_{IJ} y^J_{0} + e_{IJ}^{a}y^J_{a} + b_I(x, y)=0
\fl{Original}$$
which is not symmetric hyperbolic, one can write it as
$$
e_{IJ} y^J_{0} + e_{[IJ]}^{a}y^J_{a} + e_{(IJ)}^{a}y^J_{a} + b_I(x, y)=0
\fn$$
Now instead of looking for solutions for this operator let us split it and look for solutions of the system
$$
\cases{
&e_{IJ} y^J_{0} + e_{(IJ)}^{a}y^J_{a} + b_I(x, y)=0\cr
& e_{[IJ]}^{a}y^J_{a}=0\cr
}
\fl{SplitSystem}$$

Now one can proceed in at least two ways.

\NewSubSection{The covariant symmetrization}

Depending on the fields which are defined on $\calC$ one can look for a tail $q_I(t, x, y)$ to add which makes the two equations in the system \ShowLabel{SplitSystem}
covariant.
If such a tail  exists, then the system
$$
\cases{
&e_{IJ} y^J_{0} + e_{(IJ)}^{a}y^J_{a} + b_I(x, y)- q_I=0\cr
& e_{[IJ]}^{a}y^J_{a} + q_I=0\cr
}
\fl{CovSplitSystem}$$
is again covariant. The second equation of \ShowLabel{CovSplitSystem} is a constraint on the Cauchy surface $S$; since it does not contain time derivatives, it constrains
allowed initial conditions. One can hope that the second equation is elliptic and always has solutions.
For any solution of the second equation of  \ShowLabel{CovSplitSystem} one can define a Cauchy problem for the first equation of \ShowLabel{CovSplitSystem}, which is now symmetric hyperbolic and hence determines a unique evolution of fields.

Such an evolution of fields is a solution of the initial equation \ShowLabel{Original} with the imposed initial condition. 

\Note
\New{A similar procedure has been considered at the level of field equation in \ref{Ringstrom} as well as at the level of the action function in \ref{FF}. 
However, in  both cases they fix a background connection in order to obtain covariant theory. 
Background fields have been longly discussed to spoil the geometric essence of general relativity.
Here we instead mean to obtaining covariant equations using the fields that are already in use without adding backgrounds.
For example, doing Palatini formalism one can use the independent connection to define a non-trivial covariant derivative of the metric tensor.
}
\endNote

\NewSubSection{The non-covariant symmetrization}

If one cannot define a suitable tail to return to covariant equations, then precisely because these are not covariant one can look for a coordinate system in which the second equation of \ShowLabel{SplitSystem} is satisfied (as done in standard GR by going to harmonic coordinates).
If this is always possible (i.e.~if there exists a coordinate system for any configuration so that the second equation is satisfied)
then the first equation of \ShowLabel{SplitSystem} determines a unique evolution for fields.

The operator associated to the first equation of \ShowLabel{SplitSystem} is not the original operator any longer (due to the subtraction of the antisymmetric part of the principal symbol).
However, a solution of the symmetrized operator (i.e.~the first equation in \ShowLabel{SplitSystem}) in a coordinate system in which also the second equation of \ShowLabel{SplitSystem} is satisfied, is a solution of the original problem \ShowLabel{Original}.
Finally, being the original operator covariant, knowing a solution for it in a particular coordinate system gives as a solution of \ShowLabel{Original} in any coordinate system.

 \ms
 Let us stress that in both cases, strictly speaking, one does not solve in fact the original equation \ShowLabel{Original}.
 
In the covariant symmetrization scheme a covariant constraint is added to the theory.
Let us stress however, that we do not used any notion similar to a Poisson structure to determine them (which of course would be non-canonical in Lagrangian formalism).

 In the non-covariant symmetrization scheme the (non-covariant) equation which is in fact solved  (i.e.~the first equation in \ShowLabel{SplitSystem}) coincides with the original \ShowLabel{Original} only in a family of coordinate systems, while it differs from it in general.
If one finds a solution for standard GR and then goes in a non-harmonic coordinate system, the new metric is a solution of Einstein equations though it does not solve the non-covariant symmetrized Cauchy problem.

\NewSection{Examples}

Let us consider some examples.

\NewSubSection{Electromagnetism}

Let us consider Maxwell electromagnetism on Minkowski spacatime $(\R^4, \eta)$ as an example.
Let us fix Cartesian coordinates on spacetime (so that the metric is $\eta=\diag(-+++)$) and the ADM foliation $\pi:\R^4\arr\R: x^\mu\mapsto x^0$
so that the space hypersurfaces  are given by $S_t=\{x\in \R^4: x^0=t\}$.

The configuration bundle accounts for the quadripotential $A_\mu$.
In other words $\calC$ has fibered coordinates $(x^\mu, A_\mu)$. 
Fields can be adapted to the ADM foliation by defining
$$
A= A_0
\qquad\qquad
\vec A= A_i
\fn$$
which are a scalar field and a vector field on $S=\R^3$, respectively.

Electromagnetism is a first order theory and the one can define the field strength
$$
F_{\mu\nu}:=d_\mu A_\nu - d_\nu A_\mu
\qquad\qquad
F^{\mu\nu}:=\eta^{\mu\al}\eta^{\nu\be} F_{\al\be} 
\fn$$
Also the field strength can be projected to space by setting
$$
F_{0i}= d_0 A_i - d_i A_0
\qquad\qquad
F_{ij}= d_i A_j - d_j A_i
\quad\then
\cases{
&B^k := \frac[1/2]\ep^{ijk} F_{ij}=  \(\curl (\vec A)\)^k\cr
& E^i:= F^{0i}\cr
}
\fn$$

The dynamics is induced by the Maxwell Lagrangian
$$
L=-\frac[1/4] F_{\mu\nu}F^{\mu\nu} =  \frac[1/2]\( |E|^2 - |B|^2\)
\fn$$
where we used the fact that $F_{ij}= \ep^k{}_{ij} B_k$.

Field equations are
$$
d_\mu F^{\mu\nu}=0
\fn$$
which can be projected to space as
$$
\cases{
& d_i F^{i0}=0\cr
& d_0 F^{0i}+ d_j F^{ji}=0\cr
}
\qquad\then
\cases{
& \na \cdot E=0\cr
& d_0 E - (\na\times B)=0\cr
}
\fl{ME}$$
The first equation is clearly a constraint equation, while the second is a candidate for evolution equation.

The evolution equation reads as
$$
d_0 E^i - \ep^{ijk} d_j B_k=0
\fn$$
which are three equations for the four fields $A_\mu$.
Accordingly, one of the fundamental fields can be chosen at will in view of gauge freedom, $A'_\mu= A_\mu + d_\mu \al$.
One can show that for any $A_\mu$ one can always choose $\al$ so that $A_0=0$, which is called the {\it temporal gauge}.
The component $A_0$ is essentially left arbitrary by equations and fortunately can be fixed to zero by the gauge symmetry.
Then the physical fields are $A_i$.
In terms of the physical fields one has
$$
\cases{
&B^k := \frac[1/2]\ep^{ijk} F_{ij}= \ep^{kij} d_i A_j \cr
& E^i:= F^{0i}= -d_0 A_i \cr
}
\fl{EBFields}$$
and the evolution field equation becomes 
$$
\de^{im}\del_{00} A_m + (\de^{i(l} \de^{j)m} - \de^{im}\de^{jl})  \del_{jl} A_m=0
\fn$$
Then one has the leading coefficients
$$
e^{ij}= \de^{ij}
\qquad\qquad
e^{im jl}= (\de^{i(l} \de^{j)m} - \de^{im}\de^{jl}) 
\fn$$
The first coefficient $e^{ij}= \de^{ij}$ is symmetric, non-degenerate and \New{positive definite}.
It is a strictly Riemannian metric on the space of physical fields $A_i$.

The second coefficient $e^{im jl}$ is symmetric in $(im)$. 
The evolution equation is symmetric hyperbolic and given initial conditions determines the vector potential uniquely.
The vector potential defines electric and magnetic field by \ShowLabel{EBFields}.

\Note
The other Maxwell equations are identically satisfied in view of \ShowLabel{EBFields}.

In fact, one has
$$
\na\cdot B =   \ep^{kij} d_{ik} A_j=0
\fn$$
and
$$
\del_0 B + \na\times E= \ep^{kij} d_{i0} A_j - \ep^{kij} d_{0i} A_j=0
\fn$$
\endNote

\NewSubSection{Maxwell equations} 

By considering Maxwell equations in vacuum as equations for the electric and magnetic fields, one obtains Maxwell equations in the form \ShowLabel{ME}, i.e.
$$
\cases{
& \del_i E^i=0\cr
& \del_i B^i=0\cr
}
\qquad\qquad
\cases{
& \de_{ik} d_0 E^k + \ep_{ik}{}^{j}\del_j B^k =0\cr
& \de_{ik} d_0 B^k - \ep_{ik}{}^{j}\del_j E^k =0\cr
}
\fn$$ 
The first two equations are constraints and the second pair are first order evolutionary equations for the fields $E^A=(E^k, B^k)$.
 The leading terms are
 $$
 e_{AB}=\(\matrix{
 \de_{ik}&0 \cr
 0 &\de_{ik} \cr
 }\)
 \qquad\qquad
e^j_{AB}=\(\matrix{
 0 & \ep_{ik}{}^{j} \cr
 -\ep_{ik}{}^{j} &0 \cr
 }\) \fn$$
The time coefficients $e_{AB}$ is symmetric, non-degenerate, \New{positive definite} form.
The space coefficients $e^j_{AB}$ are symmetric. Thus the evolution equations are symmetric hyperbolic and determine the fields.
The second constraint $\na\cdot B=0$ implies that there exists a vector $\vec A$ such that $B= \na\times \vec A$.
At that point the second evolutionary equation imples that $\vec E= -\del_0 \vec A$.
Accordingly, the previous variational setting is completely recovered.

\NewSubSection{Covariant standard GR}

As shown in \ref{Simon1} the evolutionary part of the equation for standard GR is
$$
A_{\{ij\} \{lm\}} \del_{0} \del_{0} \ga^{lm} + B_{\{ij\}\{mn\}}^{kl} \del_{k} \del_{l} \ga^{mn} \approx 0
\fn$$
where $\approx$ means equality modulo a lower order tail.
Here the coefficients are
$$
A_{\{ij\} \{lm\}} =  \ga_{i(l} \ga_{m) j}
\qquad
B_{\{ij\}\{mn\}}^{kl} = -\ga^{kl}\ \ga_{i(m} \ga_{n) j} - \ga_{mn} \de_{i}^{(k}\de_{j}^{l)} 
+ \ga_{j(n} \de^{(k}_{ m)} \de_{i}^{l)} + \ga_{i(m} \de_{ n)}^{(k}\de_{j}^{l)}
\fn$$
and the spacetime metric has been decomposed as
$$
g_{\mu \nu} = \left(
\matrix{
-1 & 0 \cr
0 & \ga_{ij}
}
\right)
\fn$$

The leading time coefficient $A_{\{ij\} \{lm\}}$ is symmetric, non-degenerate and \New{positive definite}  as required.
 The leading space coefficient though is not symmetric. Its antisymmetric part is
 $$
B^{kl}_{[\{ij\} \{lm\}]} = \frac[1/2] \( \ga_{mn} \de^{(k}_i  \de^{l)}_j - \ga_{ij} \de^{(k}_m  \de^{l)}_n \).
 \fn$$
 This could be made to vanish in particular coordinate systems (e.g.~harmonic coordinate but also more general coordinates; see \ref{Simon1}).
 However, if one has, for some reason, a connection $\bar \Ga$ on spacetime it can be turned to a covariant constraint
 $$
 \frac[1/2] \( \ga_{mn} \de^{(k}_i  \de^{l)}_j - \ga_{ij} \de^{(k}_m  \de^{l)}_n \) \bar\na_{kl} \ga^{mn} =0
 \fn$$
 which in fact does not depend on time derivatives of the field $\ga^{mn}$; see \ref{Ringstrom}.

\NewSection{Quantization}

Let us here sketch a quantization scheme for field theories which can be treated in the framework introduced above.
The quantization scheme has been proposed for standard GR by Rovelli; see \ref{RovelliBook}.

First of all let us consider the boundary $\Si$ of a $m$-region $\Om\subset M$ in the form $\Om=I\times U$ with $I=[t_0, t_1]\subset \R$ and $U\subset S$.
As a special case one could consider compact close $S$ (without boundary) and $\Om=I\times S$. In this second case there is no side boundary surface $I\times \del S$  and $\Si=\Si_1- \Si_0$.

One can argue that  quantities that can be measured in quantum physics are only probabilities of boundary values of fields on $\Si$.
For this reason one can define a {\it prequantum configurations} to be the collection of the values of fields on such a boundary $\Si=\del \Om$.
Let $i:\Si\arr M$ be the canonical embedding and $\calC_\Si=i^\ast(\calC)$ be the configuration bundle restricted to $\Si$.
Prequantum configurations are sections of $\calC_\Si$ which in fact provide candidates for values of fields on $\Si$.

Given a prequantum configuration $\psi:\Si\arr \calC_\Si$ one can ask whether there exists a solution of field equations on $M$ which restricted to $\Si$
agrees with the values dictated by $\psi$. In other word, one can ask whether the prequantum configuration $\psi$ can be extended to $M$ by a solution of classical field equations. If that happens $\psi$ is called a {\it prequantum state}.

Now $\Om$ is a portion of spacetime which is globally hyperbolic by construction. If $I=[t_0, t_1]$ is an interval, then the boundary of $\Om$ is the union of three surfaces,
namely $U_0={t_0}\times U$ which is called the {\it initial surface}, $U_1={t_1}\times U$ which is called the {\it final surface}, and $S=I\times \del U$
which is called the {\it flow surface}.
Any quantum experiment can be modelled by preparing a specific configuration of fields on $U_0$, controlling what enters and exits the experimental region through $S$ and measuring the output of the experiment by reading the field values on $U_1$.
Given a prequantum state $\psi$ is measuring which classical output can be observed in this experimental setting. 

Classical field equations on $M$ do determine equations for prequantum states (which are the classical limit of a quantum description of the system).
Field equations on $M$ can in fact be split into evolution-like equations (which involve maximal order $k$ time derivatives) and constraints-like equations
(which involve time derivatives only up to order $k-1$).

Then if one finds a prequantum state on the boundary $\Si$ there is  a solution of field equations which induces that fields on the boundary.

Quantization scheme is associated to defining a Hilbert space to prequantum configurations, defining a quantum analogous of constraint equations in order to constraint functionals of the prequantum configurations.
Then restricting the Hilbert space to a Hilbert space called  the {\it physical space} which corresponds to solutions of quantum constraints.
Finally, one can define operators on the physical space which define quantum observables.   

Let us stress the different attitude between the classical and quantum setting.
When interested to classical solutions one starts from a solution of constraints to determine a classical solution. When interested in
quantization, one instead considers constraints as the real equation to be quantized while the evolution part of classical equation is used to propagates classical limits which contains information (the values of fields inside the boundary) which are not physically accessible from a quantum viewpoint. 
 
The application of this framework to LQG can be found in \ref{RovelliBook}.

 \ 
 
\NewSection{Conclusions and perspectives}

We showed that one can obtain the structure of Dirac-Bergman constraints by working within Lagrangian framework only, without resorting to Hamilton formalism, by requiring that the evolution problem is symmetric hyperbolic.

Here we overlooked that in Euler-Lagrange operators one starts with as many field equations as fundamental fields in configuration bundles.
Hence when the Lagrangian is degenerate and only some of field equations are evolution equations (while the other take the form of constraints)
then one has less evolution equation than fields. Thus the fields themselves need to be split into fields which are determined by evolution and a number of fields (which are as many as the number of constraints) which are not determined by evolution and can be fixed at will.
In specific examples this can be seen explicitly; for example in standard GR one sees directly that lapse and shift are not determined by equations and correspond to freedom to choose the ADM fibration. 
Further investigations are needed to consider the situation in general and obtain a more algorithmic procedure for splitting the fields into evolutionary degrees of freedom and gauge fields.

All information are algebraically encoded in the principal symbol of Euler-Lagrange equations.
Let us also remark that our treatment of principal symbol is better than the standard analytical approach since it does not assume a linear or affine structure on configuration bundle and  for Euler-Lagrange operators the definition of principal symbol is canonical and does not rely on a metric structure.  
Future investigations need to be devoted to characterising Lagrangians for which Euler-Lagrange equations splits in a symmetric hyperbolic evolution equation and an elliptic  system of constraints. 
One should consider if Euler-Lagrange operators for Lagrangians which can be characterized within this framework can be splitted canonically and if the analysis also teaches how fields should be split and adapted to the constraints in order to manifestly render the splitting between constraints and evolution equations. 

Also a better relation should be established between Lagrangian constraints and Hamiltonian Dirac-Bergman treatment of constraints is needed.
More examples of analysis of constraints of covariant theories is needed in the first place.

\NewAppendix{A}{Cauchy theorem for second order operators}

Let us briefly review hereafter how one can obtain Cauchy theorem for second order operators, from quasi-linear first order symmetric hyperbolic ones.
The procedure is interesting since it shows how extra conditions on the principal symbol appear at higher orders besides symmetry.

As done in \ShowLabel{SOO} we define a quasi-linear second order system as follows:
$$
\matrix{
\al_{IJ}(x, y) \del_{00} y^J - \al_{IJ}^i(x, y) \del_{0i} y^J -  \al_{IJ}^{ij}(x, y) \del_{ij} y^J 
+ \ga_I(x, y, d y) = 0.
}
\fl{secondOrder}
$$
and its associated Cauchy problem:
$$
\cases{
&\al_{IJ} \del_{00} y^J - \al_{IJ}^i \del_{0i} y^J -  \al_{IJ}^{ij} \del_{ij} y^J + \ga_I= 0. \cr
&\qquad y^J(0, x^i) = f^J(x^i) ,
\quad \del_0 y^J(0, x^i) = g^J(x^i).
}
\fl{secondOrderCP}
$$
which will be called {\it CP2}.

Our goal is to transform a second order PDE in a first order system, by introducing auxiliary fields. 
Inspired by the method used for ODE we can define the following new fields:
$$
\cases{
&v^J = \del_0 y^J \cr
&v^J_j = \del_j y^J \cr
}
\qquad\then
\del_0 v^J_i = \del_i v^J
\fn$$
However, if we wish to consider these equations as part of the original system one should notice that they do not correspond to differential operators with values in the dual space of field variations as it was for the original equation.
Then one should introduce some suitable bilinear forms to write them equivalently in the form:
$$
\cases{
&\be_{IJ} \(\del_0 y^J - v^J \)= 0 \cr
&\be_{IJ}^{ij} \( \del_0 v^J_j - \del_j v^J \) = 0
}
\qquad\then \be_{IJ}^{ij}\(v^J_j - \del_j y^J\)=0 
\fn$$
for some invertible coefficients $\be_{IJ}$ and $\be_{IJ}^{ij}$.  

Let us remark that the equation $v^J_j = \del_j y^J $ contains no time derivative and as such is a constraint on initial conditions and will not contribute to the Cauchy problem.

Then the equation  \ShowLabel{secondOrder}  can be written in terms of the new fields $(y^I, v^I, v^I_i)$, so that its Cauchy problem can be recast in the following form:
$$
\cases{
&\be_{IJ} \del_0 y^J \approx 0\cr
&\al_{IJ} \del_{0} v^J - \al_{IJ}^i \del_i v^J - \al_{IJ}^{ij} \del_{i} v_j^J \approx 0\cr
&\be_{IJ}^{ij} \del_0 v^J_j - \be_{IJ}^{ij} \del_j v^J  = 0\cr
&\qquad y^J(0, x^i) = f^J(x^i) ,\quad
 v^J(0, x^i) = g^J(x^i), \quad
 v^J_i(0, x^i) = \del_i f^J(x^i)
}
\fl{firstOrderReduced}$$
together with the constraint $\del_i y^J = v_i^J$.
We can write this system in the block-matrix form as follows:
$$
\(
\matrix{
\be_{IJ} &        0       &            0 \cr
    0         &   \al_{IJ} &            0 \cr
    0         &         0       &   \be_{IJ}^{ij} 
}
\)
\del_0 \(
\matrix{
y^J \cr
v^J \cr
v^J_j
}
\)
-
\(
\matrix{
    0         &           0              &           0               \cr
    0         &    \al^k_{IJ}        &   \al_{IJ}^{kj}     \cr
    0         &    \be_{IJ}^{ik}  &    0
}
\)
\del_k \(
\matrix{
y^J \cr
v^J \cr
v^J_j
}
\)
\approx 0
\fl{Eq1}$$

This is a first order CP so Cauchy theorem applies to it. One has existence and uniqueness of solutions if
the first matrix is symmetric, non-degenerate,  positive definite  and the second is symmetric.

We already know that $\al_{IJ}$ is non-degenerate and positive definite. 
If also $\be_{IJ}$ and $\be_{IJ}^{ij} $ are non-degenerate and positive definite then the whole matrix is.
Since we are free to choose $\be_{IJ}$ as we wish (provided that the choice is non-degenerate and positive definite)
we can fix it as $\be_{IJ}= \al_{IJ}$, which is automatically a good choice.

For the second matrix to be symmetric $\al^k_{IJ}$ must be symmetric and one must have
$$
\be_{IJ}^{ij} =  \al_{JI}^{ij}
\fl{SymmetricMatrix}$$
Thus the block $\be_{IJ}^{ji}$ (and as a consequence of this choice the coefficient $\al_{IJ}^{ij}$)  must be symmetric in $(IJ)$
and non-degenerate positive definite.

We can thus rewrite the original system as:
$$
\cases{&
\(
\matrix{
\al_{IJ} &        0       &            0 \cr
    0         &   \al_{IJ} &            0 \cr
    0         &         0       &   \al_{IJ}^{ij} 
}
\)
\del_0 \(
\matrix{
y^J \cr
v^J \cr
v^J_j
}
\)
-
\(
\matrix{
    0         &           0              &           0               \cr
    0         &    \al^k_{IJ}        &   \al_{IJ}^{kj}     \cr
    0         &    \al_{IJ}^{ki}  &    0
}
\)
\del_k \(
\matrix{
y^J \cr
v^J \cr
v^J_j
}
\)
\approx 0\cr
&\qquad y^J(0, x^i) = f^J(x^i) ,\quad
 v^J(0, x^i) = g^J(x^i), \quad
 v^J_i(0, x^i) = \ h^J_i(x^i)
}
\fl{RS}$$
which, together with the constraint $\del_i y^J = v_i^J$,  is called the {\it reduced Cauchy problem} or {\it CP1} for short.

Let us remark that once again, also for second order operators, the well-posedness of CP2 is subjected to algebraic requirements.
Unlike for first order operators symmetry is no longer enough and one needs to require that  the coefficient $\al_{IJ}^{kj}$ is also positive definite.

Finally, one should show that {\it CP1} and {\it CP2} are dynamically equivalent, i.e.~there exists a 1--to--1 correspondence of solutions, 
so that well-posedness of {\it CP1} implies  well-posedness  {\it CP2}.
For that we refer to \ref{Simon1}.

\Acknowledgements

We acknowledge the contribution of INFN (Iniziativa Specifica QGSKY), 
the local research project {\it Metodi Geometrici in Fisica Matematica e Applicazioni} (2015) of Dipartimento di Matematica of University of Torino (Italy). 
This paper is also supported by INdAM-GNFM.

We are grateful to M.Ferraris and C.Rovelli for discussions and comments.

\ShowBiblio

\end